\documentclass{article}
\usepackage{amssymb}
\def\ov{\overline}
\def\ds{\displaystyle}

\def\res{\mathop{\mathrm {res}}\limits_}

\makeatletter
\@addtoreset{equation}{section}

\newtheorem{theorem}{Theorem}[section]
\newtheorem{examp}{Example}[section]
\newtheorem{coroll}{Corollary}[section]
\newtheorem{examps}{Examples}[section]

\newtheorem{lemma}{Lemma}[section]
\newtheorem{remark}{Remark}[section]

\newtheorem{remarks}[remark]{Remarks}
\newtheorem{proposition}{Proposition}[section]
\newtheorem{definition}{Definition}[section]

\def\le{\left}
\def\m{\mathop}
 \def\tr{{\rm Tr}}

\def\ri{\right}
\def\br{\begin{remark}\rm\small}
\def\1{{\bf 1}}
\def\er{\end{remark}}
\def\bt{\begin{theorem}\rm}
\def\et{\end{theorem}}
\def\bc{\begin{coroll}\rm}
\def\ec{\end{coroll}}
\def\brs{\begin{remarks}.\\ \rm\small\begin{enumerate}}
\def\ers{\end{enumerate}\end{remarks}}
\def\bx{\begin{examp}\small}
\def\ex{\end{examp}}
\def\bl{\begin{lemma}\small}
\def\el{\end{lemma}}
\def\bxs{\begin{examps}. \rm\begin{enumerate}}
\def\exs{\end{enumerate}\end{examps}}
\def\bd{\begin{definition}}
\def\ed{\end{definition}}
\def\bp{\begin{proposition}\rm}
\def\ep{\end{proposition}}
\def\be{\begin{equation}}
\def\ee{\end{equation}}

\def\bea{\begin{eqnarray}}
\def\eea{\end{eqnarray}}
\def\beas{\begin{eqnarray*}}
\def\eeas{\end{eqnarray*}}
\def \pa{\partial}
\def\C{{\mathbb C}}

\def\R{{\mathbb R}}
\def\N{{\mathbb N}}

\def\wt{\widetilde}
\begin{document}
\begin{flushright}
CRM-2931 (2003)
\end{flushright}
\vspace{0.2cm}
\begin{center}
\begin{Large}
\textbf{Second and  Third Order Observables of the  Two--Matrix
  Model\footnote{
Work supported in part by the Natural Sciences and Engineering Research Council
of Canada (NSERC), Grant. no. 261229-03.}}
\end{Large}\\
\vspace{1.0cm}
\begin{large} {M.
Bertola}$^{\dagger\ddagger}$\footnote{bertola@mathstat.concordia.ca}
\end{large}
\\
\bigskip
\begin{small}
$^{\dagger}$ {\em Department of Mathematics and
Statistics, Concordia University\\ 7141 Sherbrooke W., Montr\'eal, Qu\'ebec,
Canada H4B 1R6} \\ 
\smallskip
$^{\ddagger}$ {\em Centre de recherches math\'ematiques,
Universit\'e de Montr\'eal\\ C.~P.~6128, succ. centre ville, Montr\'eal,
Qu\'ebec, Canada H3C 3J7} \\
\end{small}
\bigskip
\bigskip
%%%%%%%%%%%%%%%%%%%%%%%%%%%%%%%%%%  Abstract %%%%%%%%%%%%%%%%%%%%%%%%%%%%%%%
{\bf Abstract}
\end{center} 
\begin{center}
\begin{small}
\parbox{13cm}{
In this paper we complement our recent result on the explicit formula for the  planar
limit of the free energy of the two--matrix model by computing the
second and third order observables of the model in terms of canonical
structures of the underlying genus $g$ spectral 
curve.  In particular we provide explicit formulas for  any
three--loop correlator of the model. Some explicit examples are
worked out.
}
\end{small}
\end{center}
%%%%%%%%%%%%%%%%%%%%%%%%%%%%%%%%%%%%%%%%%%%%%%%%%%%%%%%%%%%%%%%%%%%%%%%%%%%%%%%%
%
%                                                                               
%
%                       Introduction begins...                                   
%
%                                                                               
%
%                                                                               
%
%%%%%%%%%%%%%%%%%%%%%%%%%%%%%%%%%%%%%%%%%%%%%%%%%%%%%%%%%%%%%%%%%%%%%%%%%%%%%%%%
%
\section{Introduction}
The present paper is the natural complement of our recent publication
\cite{F1} in which an explicit residue formula for the free-energy in
the planar limit (and for arbitrary genus spectral curves) of
the Hermitian two--matrix model was derived. In point of fact the
present results, while logically consequent that paper, can be (and in
fact are) derived without using the explicit formula in \cite{F1}.
Given the well-known relation of the free--energy of this model with
the tau--function of the dispersionless Toda hierarchy, we could as
well rephrase the entire contents of this (and the previous) paper in
the language of integrable hierarchies.\par
Let us briefly recall the historical and logical  setting of the problem. 
The object under scrutiny is the
random 2-matrix model \cite{Mehta, BI}, which has 
many applications to  solid state
physics \cite{Guhr} (e.g., conduction in mesoscopic devices,  quantum chaos and,
lately, crystal growth\cite{spohn}), particle  physics \cite{Verbaarshot},
$2d$-quantum gravity and string theory \cite{Matrixsurf, ZJDFG, DVV}.
The model consists of two Hermitian matrices
$M_1,M_2$ of size $N\times N$ with a probability distribution given by
the formula 
\bea
{\rm d}\mu(M_1,M_2) = \frac 1{\mathcal Z_N} {\rm d}M_1{\rm d}M_2
\exp\le[-\frac 1 \hbar\tr  \le(V_1(M_1)+V_2(M_2)-M_1M_2\ri)\ri]\
,\nonumber \\
V_1(x) = \sum_{K=1}^\infty \frac {u_K}K x^K\ ;\qquad 
V_2(y) = \sum_{J=1}^{\infty}\frac {v_J}J y^J\ ,
\label{measure}
\eea
where the functions $V_i$ (called {\em potentials}) are polynomials of
degree $d_i+1$ for simplicity (and definiteness), but could be
formally extended to power series.  
The partition function $\mathcal Z_N$ is known to be a $\tau$-function
for the KP hierarchy in each set of deformation parameters
(coefficients of $V_1$ or $V_2$) and to provide solutions of the
two--Toda hierarchy \cite{UT, AvM1, AvM2}.
This model with polynomial potentials has been investigated in the series of paper
\cite{BHE1, BHE2, BHE3} where a duality of spectral curves and
differential systems for the relevant biorthogonal polynomials has
been unveiled and analyzed in the case of polynomial potentials. In
\cite{BE} the mixed correlation functions of the model (traces of
powers of the two non-commuting matrices) have been reduced to a
formal Fredholm-like determinant without any assumption on the nature
of the potentials and using the recursion coefficients for the
biorthogonal polynomials.
We  recall that the biorthogonal polynomials are two sequences
of monic polynomials  (\cite{BHE1} and references therein)
\be
\pi_n(x) = x^n + \cdots , \qquad \sigma_n(y)=y^n + \cdots, \qquad n=0,1,\dots
\ee
that are ``orthogonal'' (better say ``dual'') w.r.t. to the coupled
measure  on the product space
\be
\int_\R\!\!\int_\R\!\!\! {\rm d}x\, {\rm d}y \,\, \pi_n(x)\sigma_m(y) {\rm
e}^{- \frac 1\hbar(V_1(x)+ V_2(y) -xy)} = h_n\delta_{mn} ,\qquad h_n\neq 0\ \forall n\in\N\label{norms}
\ee
where $V_1(x)$ and $V_2(y)$ are  the functions in the  measure (\ref{measure}).
It is convenient to introduce the associated quasipolynomials
 defined by the formulas
\bea
&&\psi_n(x):= \frac 1{\sqrt{h_{n-1}}}\pi_{n-1}(x){\rm e}^{-\frac 1
  \hbar V_1(x)}\\
&&\phi_n(y):= \frac 1{\sqrt{h_{n-1}}}\sigma_{n-1}(y){\rm e}^{-\frac 1
  \hbar V_2(y)}\ .\label{quasidifferentials}
\eea
In terms of these two sequences of quasipolynomials the multiplications
by $x$ and $y$ respectively are represented by semiinfinite square
matrices $Q = [Q_{ij}]_{i,j\in \mathbb N^*}$ and  $P = [P_{ij}]_{i,j\in
  \mathbb N^*}$ according to the formulae
\bea
&&x\psi_n(x) = \sum_m Q_{n,m}\psi_m(x)\ ;\ \ \ y\phi_n(y) = \sum_m
P_{m,n}\phi_m(y) \nonumber \\
&&Q_{n,m}=0=P_{m,n}, \ {\rm if}\ n>m+1.\label{mulxy}
\eea
The matrices $P$ and $Q$ satisfy the
``string equation'' 
\be
[P,Q]=\hbar \1\ \label{string}\ .
\ee
 We refer for further details to \cite{Berto,BHE1,BHE2,BHE3} where
these models are studied especially in the case of polynomial potentials. We 
also point out that the model can easily be generalized to accommodate
contours of integration other than the real axes and arbitrary
(possibly complex) potentials \cite{Berto,BHE1}.  

The partition function is believed to have a large $N$ expansion (with
$\hbar = \frac t N$ and $t$ fixed) according to the formula
\be
-\frac 1 {N^2} \ln\mathcal Z_N = \mathcal F =  \mathcal F^{(0)} + \frac 1
{N^2} \mathcal F^{(1)} + \cdots\ .
\ee
This expansion in powers of $N^{-2}$ has been repeatedly advocated for
the $2$-matrix model on the basis of physical arguments \cite{ZJDFG,
  eynard2, Bertrand_unpublished}
and has been rigorously proven in the one-matrix model \cite{ercol}.
In the two-matrix model this expansion is believed to generate
2-dimensional statistical models of surfaces triangulated with
ribbon-graphs \cite{ZJDFG, Matrixsurf, Kazakov}, where the powers of
$N^{-1}$ are the Euler characteristics of the surfaces being
tessellated.
From this point of view the term $\mathcal F^{(0)}$ corresponds to a genus
$0$ tessellation and the next to a genus one tessellation.
Remarkably, an algorithm for the computation of the subleading terms
in the $\hbar=\frac t N $ expansion
is known and also a closed expression of the genus $1$
correction $\mathcal F^{(1)}$  \cite{eynard2,
  Bertrand_unpublished}.\par\vskip 5pt
The object of \cite{F1} was the leading term of the free energy,
$\mathcal F^{(0)}$. It is the generating function (to leading order) of the expectations of
the powers of the two matrices in the model
\be
\langle \tr {M_1}^K\rangle = K\pa_{u_K} \mathcal F^{(0)} + \mathcal
O(N^{-2})\ ,\qquad  \langle {\tr M_2}^J\rangle = J\pa_{v_J} \mathcal F^{(0)} + \mathcal
O(N^{-2})\ .
\ee
It is at the same time the logarithm of the $\tau$-function of the
dispersionless Toda hierarchy (in the one--cut case) \cite{tt1,tt2}; this hierarchy falls in a
broad sense in the same sort of integrable dispersionless hierarchies
called ``universal Whitham hierarchy'' (i.e. dispersionless KdV and
KP)  and studied in \cite{Kric},
where formulas for the corresponding $\tau$-function are provided
which are quite similar to that in \cite{F1}. However the situation
under inspection in this paper (and in \cite{F1}) involves a hierarchy
which does not quite fall in the very broad class of \cite{Kric};
there the moduli space under consideration was the set of curves of
genus $g$ with $N$ marked points $\infty_k$, $k=1,\dots,N$ and local parameters around the
punctures, together with a fixed meromorphic differential of the
second kind ${\rm d}\mathcal E$ with poles of degree $n_k+1$ at
$\infty_k$.
The (local) dimension of the space of these data is
$3g-2+\sum_{k=1}^{N}(n_k+1)$. \par\vskip 4pt
The object of interest  of the present paper is -among others- the
explicit computation of the  double and triple
correlators
\bea
\le\langle\tr {M_i}^a \tr{M_j}^b \ri\rangle_{conn}\ ,\qquad
\le\langle\tr {M_i}^a \tr{M_j}^b \tr{M_k}^c\ri\rangle_{conn}\ , \ \
i,j,k=1,2, \ a,b,c\in \N.
\eea
or --which is the same-- their generating functions 
\bea
&&\le\langle\tr \frac 1 {x_1-M_1} \tr \frac
1{x_2-M_1}\ri\rangle_{conn} = \frac \delta{\delta V_1(x_1)}  \frac 
\delta{\delta V_1(x_2)}\mathcal F\cr
&& \le\langle\tr \frac 1 {x-M_1} \tr \frac 1{y-M_2}\ri\rangle_{conn} =
\frac \delta{\delta V_1(x)}  \frac 
\delta{\delta V_2(y)}\mathcal F\cr
&&\le\langle\tr \frac 1 {y_1-M_2} \tr \frac
1{y_2-M_2}\ri\rangle_{conn} = \frac \delta{\delta V_2(y_1)}  \frac 
\delta{\delta V_2(y_2)}\mathcal F\ ,\label{2loop}
\eea
\bea
\le\langle\tr \frac 1 {x_1-M_1} \tr \frac 1{x_2-M_1} \tr \frac 1 {x_3
  - M_1} \ri\rangle_{conn} = \frac \delta{\delta V_1(x_1)}  \frac
\delta{\delta V_1(x_2)} \frac \delta{\delta V_1(x_3)}\mathcal F\cr
\le\langle\tr \frac 1 {x_1-M_1} \tr \frac 1{x_2-M_1} \tr \frac 1 {y
  - M_2} \ri\rangle_{conn} = \frac \delta{\delta V_1(x_1)}  \frac
\delta{\delta V_1(x_2)} \frac \delta{\delta V_2(y)}\mathcal F\cr
\le\langle\tr \frac 1 {y_1-M_2} \tr \frac 1{y_2-M_2} \tr \frac 1 {x
  - M_1} \ri\rangle_{conn} = \frac \delta{\delta V_2(y_1)}  \frac
\delta{\delta V_2(y_2)} \frac \delta{\delta V_1(x)}\mathcal F\cr
\le\langle\tr \frac 1 {y_1-M_2} \tr \frac 1{y_2-M_2} \tr \frac 1 {y_3
  - M_2} \ri\rangle_{conn} = \frac \delta{\delta V_2(y_1)}  \frac
\delta{\delta V_2(y_2)} \frac \delta{\delta V_2(y_3)}\mathcal F\ ,\label{3loop}
\eea 
where the ``puncture'' operators are defined (a bit formally) as  
\be
\frac \delta{\delta V_1(x)} := \sum_{K=1}^\infty x^{-K-1} K \pa_{u_K}\
,\qquad 
 \frac \delta{\delta V_2(y)} := \sum_{J=1}^\infty y^{-J-1} J \pa_{v_J}.
\ee
The meaning of the puncture operators when the potentials are
polynomials is given by the understanding that each derivative
-say- w.r.t. $u_K$  gives a $\tr {M_1}^K$ contribution inside the
expectation value. The two--loops correlators (\ref{2loop}) are well
studied and our results here are not new at least in the one--cut case
\cite{daul}, but are needed for computing
the three--loop correlator (\ref{3loop}). The latter have not been
computed for  this model; the similar problem for the normal matrix
model has been addressed in \cite{WZ2} in a  setting that is a
subsetting of our, primarily due to the fact their case corresponds to
a spectral curve of genus $0$ plus certain reality conditions.
Indeed the planar limit of the free energy $\mathcal F^{(0)}$ (we will drop
the superscripts from now on) can be defined for any spectral curve
$\Sigma_g$ of genus $g$ as a closed differential on a certain moduli
space and this differential can be integrated explicitly (\cite{F1}
and references therein). In the literature on matrix models the genus
of this spectral curve is related to the number of connected
components  (called ``cuts'') of the support of the densities of the spectra of the two
matrices; in this terminology the case of genus $0$ spectral curves
corresponds to the {\em one cut} assumption, whereas the higher genus
case to multicut solutions.

In the present paper  we complement our previous paper \cite{F1} by
computing explicitly the second and third order derivatives of the
free energy.  As we have anticipated, however, these result are
derived independently of \cite{F1}, only the notation being borrowed
thence. We also point out that the resulting formulas are very close
to similar residue formulas in \cite{Kric} which, however, deals with
a different hierarchy and moduli space and whose proofs are not given
explicitly therein.\par\vskip 6pt
The paper is organized as follows: in Section \ref{setting} we set up
the problem and notation, recalling the formula \cite{F1} for the free energy
$\mathcal F_g$ over a spectral curve of genus $g$. We also link our
present results to the current relevant literature.
In Section \ref{bergman} we give a to-the-bone review of the Bergman
kernel on algebraic curves and of the properties that will be used in
the sequel.
In Section \ref{second} we compute all second order observables, which
include the derivatives with respect to the filling fraction and the
``temperature'' $t$ (which can be interpreted as number operator as
well depending on the points of view).
Finally, in Section \ref{third} we compute all third order observables
by deriving a formula for the variations of the conformal structure of
the spectral curve under variation of the parameters of the problem.

\section{Setting and notations}
\label{setting} 
 As in our previous paper \cite{F1} we will  work with the following data:
a (smooth) curve $\Sigma_g$ of genus $g$ with two marked
points $\infty_Q$, $\infty_P$ and  two functions
$P$ and $Q$ which have the following pole structure;
\begin{enumerate}
\item The function $Q$ has a simple pole at $\infty_Q$ and a pole of
  degree $d_2$ at $\infty_P$.
\item The function $P$ has a simple pole at $\infty_P$ and a pole of
  degree $d_1$ at $\infty_Q$.
\end{enumerate}
From these data it would follow that $P,Q$ satisfy a polynomial
relation and hence that the curve is a plane algebraic (singular) curve, but we will not need it for our computations\footnote{
We could also generalize this setting to the case $d_1=d_2=\infty$
which would simply mean that the two functions have one simple pole
and one essential singularity; this generalization is quite
straightforward and would need only some care about convergence.}.
Moreover we will fix a symplectic basis of cycles $\{a_j,b_j\}_{j=1,\dots g}$ in
the homology of the curve.
By their definition we have 
\bea
P &=& \sum_{K=1}^{d_1+1} u_K Q^{K-1} -\frac t Q- \sum_{K=-1}^{-\infty}
K U_K Q^{-K-1}=:
V_1'(Q) -\frac t Q+ \mathcal O (Q^{-2}) 
\ ,\ \ \hbox{near $\infty_Q$}\nonumber\\ 
Q &=&\sum_{J=1}^{d_2+1} v_J P^{K-1} -\frac t P- \sum_{J=-1}^{-\infty}
JV_JP^{-J-1} =: V_2'(P) -\frac t P+ \mathcal O (P^{-2})
\ ,\ \ \hbox{near $\infty_P$}.\label{cazzata}
\eea
The fact that the coefficient of the power $Q^{-1}$ or $P^{-1}$ is the
same follows immediately from computing the sum of the residues of
$P{\rm d}Q$ (or $Q{\rm d}P$). The polynomials $V_1$ and $V_2$ defined
by the above formula (\ref{cazzata}) are the potentials of the
matrix model whose free energy we are considering\footnote{Formula (\ref{cazzata}) defines only
  their derivative; what we mean by $V_1(x)$ and $V_2(y)$ is
  explicitly
\be
V_1(x) := \sum_{K=1}^{d_1+1} \frac{ u_K}K x^K\ ,\qquad V_2(y):=
\sum_{J=1}^{d_2+1} \frac {v_J}J y^J\ .
\ee
The constant term in the two potential would have trivial consequences
as it amounts to a rescaling of the partition function of the matrix
model by ${\rm e}^{u_0+v_0}$, which of course has absolutely no
consequences on all the observables of the model.} and the functions
$Q$ and $P$ represent the semiclassical (commutative as per
eq. \ref{string}) limit of the multiplication operators for the
orthogonal polynomials.
The coefficients $u_K$, $v_J$, $t$  are read off
eqs. (\ref{cazzata}) as follows 
\be
u_K =-\res{\infty_Q} PQ^{-K}{\rm d}Q\ ,\ \ v_J
=-\res{\infty_P}QP^{-J}{\rm d}P\ ,\ \ t=\res{\infty_Q}P{\rm d}Q=
\res{\infty_P}Q{\rm d}P\ .\label{moduli}
\ee
Note that the requirement that the curve possesses two meromorphic
functions with this pole structure imposes strong constraints on the
moduli of the curve itself. In fact a Riemann-Roch argument shows that
the moduli space of these data is of dimension $d_1+d_2+3+g$; to the
above  $(d_1+1)+(d_2+1)+1$ parameters  we add the following period
integrals referred to as {\em filling fractions}
\be
\epsilon_i := \frac 1{2i\pi}\oint_{a_i}P{\rm d}Q\ ,\ \ i=1,\dots g.
\ee
Here we have introduced a symplectic basis $\{a_i,b_i\}_{i=1\dots g}$ in the homology of the
curve $\Sigma_g$ and the choice of the $a$-cycles over the $b$-cycles
is purely conventional.\par
The free energy $\mathcal F_g$ of the model (where the subscript
refers to the genus of the spectral curve, not the genus of the
$\hbar$ expansion) is then  defined by the equations
\bea
&&\pa_{u_K}\mathcal F_g = \frac 1 K \res{\infty_Q} PQ^K{\rm d}Q = U_K\ ,\qquad 
\pa_{v_J}\mathcal F_g = \frac 1 J \res{\infty_P} QP^J{\rm d}P = V_J\ ,\\
&& \pa_{\epsilon_i} \mathcal F_g = \oint_{b_j} P{\rm d}Q =:\Gamma_i.
\eea
These equations and in particular the derivatives with respect to the
filling fractions appear in this precise context in \cite{curve};
nevertheless, after identifying the free energy with some tau function
for a dispersionless Whitham hierarchy, similar formulas appear in
\cite{Kric}.

The above equations mean that the coefficients of singular parts of $P$ {\em qua} function of $Q$ (or
vice-versa) are the independent coordinates and the coefficients of the
regular part (starting from $Q^{-2}$) are the corresponding derivatives of the free
energy. Moreover the $a$-periods of $P{\rm d}Q$ are the independent
variables and the $b$-periods the corresponding derivatives of
$\mathcal F_g$.
In fact one should add the extra constraints $\Gamma_i=0$ to this functional in order
to assure that it comes from a saddle--point integration of the
two--matrix model. However, since  we are
interested here mostly with the formal $N\to \infty, \hbar\to 0,\
\hbar N = t$ limit (as explained in \cite{chaineloop}) we do not
impose the extra constraint and treat the filling fractions as
independent coordinates.\par
It was shown in \cite{F1} (but see also \cite{chaineloop} for the
extension to the chain of matrices) that $\mathcal F_g$ is given by
the following equivalent formulas
\bea
2\mathcal F_g &=&  \res{\infty_Q}P\Phi_1{\rm d}Q + \res{\infty_P}
Q\Phi_2{\rm d}P + \frac 12 \res{\infty_Q}P^2Q{\rm d}Q + t\mu +
\sum_{i=1}^{g} \epsilon_i\Gamma_i=\label{piccole}\\
&=& \sum_{K}u_K U_K +\sum_J v_J V_J +\frac 1 2 \res{\infty_Q}P^2 Q{\rm
  d}Q+t\mu +
\sum_{i=1}^{g}\epsilon_i\Gamma_i = \\
&=&\sum_K \frac{2-K}2 u_K U_K + \sum_J \frac {2-J}2 v_J V_J +t\mu +
\sum_{i=1}^g \epsilon_i\Gamma_i-\frac 1 2 t^2\ ,\\
&&\hspace{-1cm} \Phi_1(\zeta) :=\frac 1 {2i\pi}\oint_{\infty_Q} \hspace{-8pt}\ln\le(1- \frac {Q
}{Q(\zeta)}\ri)P{\rm d}Q=  -V_1(Q(\zeta))+t\ln
(Q(\zeta))+\int_{X_q}^\zeta  \hspace{-8pt}P{\rm d}Q = \sum_{K=1}^{\infty} U_K
Q^{-K}= \mathcal O (Q^{-1})\\
&&\hspace{-1cm} \Phi_2(\zeta) :=\frac 1{2i\pi} \oint_{\infty_P} \hspace{-8pt} \ln\le(1- \frac {P
}{P(\zeta)}\ri)Q{\rm d}P= -V_2(P(\zeta))+t\ln(P(\zeta)) +
\int_{X_p}^\zeta \hspace{-8pt} Q{\rm d}P = \sum_{J=1}^{\infty}V_J P^{-J}=\mathcal O (P^{-1})\\
&&\hspace{-1cm} \mu :=  \res{\infty_Q} \big[V_1(Q) -
t\ln(Q/\lambda)\big]{\rm d}S - \res{\infty_P} \big[V_2(P) -
t\ln(P \lambda)\big]{\rm d}S  -\res{\infty_Q} PQ{\rm d}S +
\sum_{i=1}^g \epsilon_i\oint_{b_i} {\rm d}S\ ,\label{grandi}
\eea
where ${\rm d}S$ is the normalized differential of the third kind with
poles at $\infty_{P,Q}$ and residues $\res{\infty_Q}{\rm d}S =- 1 =
-\res{\infty_P}{\rm d}S$ and the function
$\lambda$ is the following function (defined up to a multiplicative
constant which drops out of the above formula) on the universal covering of the
curve with a simple zero at $\infty_Q$ and a simple pole at $\infty_P$
\be
\lambda:= \exp\le(\int {\rm d}S\ri)\ . 
\ee
\br
The ``chemical potential'' $\mu$ in the context of the dispersionless
Toda hierarchy is the long--wave limit of the  Toda field \cite{tt4} (denoted by $\phi$ there)
and satisfies, among others, the Toda field equation which is written,
in our notations, 
\be
\frac {\pa^2 \mu}{\pa u_1\pa v_1} + \pa_t \exp(\pa_t\mu) = 0\ .
\ee
\er
\br
In formulas (\ref{piccole},\dots,\ref{grandi})  the open paths of integration are supposed to not
intersect the basis of cycles, i.e., to remain in the fundamental cell
of the universal covering of the curve and the residues involving the
multivalued function $\lambda$ are taken to be on the same cell of the
universal cover.
\er
We now make the necessary connections with the relevant literature
\cite{KKWZ, KMZ, MWZ1, WZ, Zab, Tak, tt1, curve}.

The equations that define here the free energy in the genus $g=0$ case
are precisely the same that define the $\tau$-function
of the dToda hierarchy where one imposes the (compatible) constraint
of the string equation to the Lax functions. In the relevant literature \cite{tt1, tt2,
  tt3, Teo} the functions $P,Q$ are the Lax operators denoted by $L, \tilde L^{-1}$ or $\mathcal
L, \tilde{\mathcal L}^{-1}$ and the normalization is slightly different.
In higher genus our free energy is related to the tau function for
solutions obtained via a Whitham averaging method on some invariant
submanifold of the hierarchy.\par
 The link with the important  works \cite{MWZ1,
 KKWZ, WZ, Zab, Tak} is  as follows; if the two potentials are complex
 conjugate $V_1= \overline{V_2} =: V$ then (in genus $0$) the
 functions  $Q$ and $P$ are conjugate-Schwarz-inverted in the 
 sense explained presently;
pick the uniformizing parameter $\lambda$  of the rational curve
 $\Sigma_0$ to have a zero at $\infty_P$ and a pole at
 $\infty_Q$(and a suitable normalization)\footnote{This $\lambda$ is exactly the same $\lambda$
 appearing in the higher genus formulas; in fact --quite obviously--  $\lambda =  \exp \int
\frac{ {\rm d}\lambda}\lambda$, which is translated in the higher
 genus setting simply by replacing $\frac {{\rm d}\lambda}\lambda$
 with the third--kind differential ${\rm d}S$.}. Then one has
\be
Q(\lambda) = \ov{ P(\ov \lambda^{-1})}
\ee
 The function $Q(\lambda)$ is then
the uniformizing map of a Jordan curve $\Gamma$ in the $Q$-plane (at
 least for suitable ranges of the parameters) which is defined by either of the following relations
\be
 \overline{Q(\lambda)}=P(1/\lambda) \ \hbox { or }\  |\lambda| = 1\ .
\ee
 In the setting of \cite{WZ, KKWZ, MWZ1}  the function $Q$ is denoted
 by $z$ (and $\lambda$ by $w$) so that  then $P$ is the Schwartz function of the curve
 $\Gamma$, defined  by 
\be
\overline z = S(z)\ ,  \ \ z\in \Gamma\ .
\ee 
The coefficients of the potential $V(x) =
\sum_{K}\frac {t_K}K x^K$ are the so-called ``exterior harmonic
moments'' of the region $\mathcal D$ enclosed by the curve $\Gamma$
\bea
&& t_K =
\frac 1 {2i\pi}\oint_{\Gamma} \overline z z^{-K}{\rm d}z\\
&& t=t_0 = \frac 1{2i\pi} \int\!\!\!\!\!\int_{\mathcal D} {\rm
  d}z\wedge{\rm d}\overline z =  \frac 1{2i\pi}\oint_{\Gamma}
\overline z {\rm d}z= \frac{Area(\mathcal D)}\pi\ .
\eea
By writing $\overline z=S(z(w))$ these integrals become residues in
the $w$-plane. For conformal maps (i.e. Jordan curves)  the $\tau$-function has been defined in \cite{Tak}
and given an appealing interpretation as (exponential of the Legendre transform of) the
electrostatic potential of a uniform $2$-dimensional distribution of charge in
$\mathcal D$ \cite{KKWZ} 
\be
\ln(\tau_\Gamma) = -\frac 1 \pi \int\!\!\!\int_{\mathcal D} \ln\le|\frac 1 z
- \frac 1 {z'} \ri|{\rm d}^2 z{\rm d}^2 z'\ .
\ee
It can be rewritten as a (formal) series in the exterior and interior
 moments as  (note that we are changing the normalization used in
 \cite{KKWZ, MWZ1, WZ, Zab} to match more closely ours)
\be
2\ln(\tau_\Gamma) = -\frac 1 {4\pi} \int\!\!\!\!\int_{\mathcal D} {\rm
 d}^2 z |z|^2 + t_0w_0 +\sum_{K>0}(t_K w_K+ \overline {t}_K \overline{w}_K)
\ee
where the interior moments are defined by (the normalization here
 differs slightly from \cite{WZ})
\bea
\pa_{t_k}\tau_\Gamma = w_K =\frac 1 {\pi K}  \int\!\!\!\!\int _{\mathcal D}z^K {\rm d}^2z\ ,\
 \ K>0\\
\pa_{t_0}\tau_\Gamma = w_0 = \frac 1{\pi}  \int\!\!\!\!\int _{\mathcal D} \ln|z|^2{\rm d}^2z\ .
\eea
The logarithmic moment $w_0$ corresponds to our ``chemical potential''
 $\mu$; it is not at all obvious (at least we failed to prove it by
 direct integration) but it is necessarily so because both 
 $\mathcal F_0$ and $\tau_\Gamma$ satisfy the same differential
 equations w.r.t. $t_K$ (or, in our notation $u_K=\ov v_K$) and the
 same scaling constraint 
\be
4\mathcal F =-t^2 + \bigg( 2t\pa_t + \sum_K (2-K)u_K\pa_{u_K} +
\sum_J(2-J) v_J\pa_{v_J}\bigg)\mathcal F\ .\label{scall}
\ee
\br
Note that  formula (\ref{piccole}) is quite effective in that it allows explicit
computations. For instance in the genus zero case one could do the
exercise (e.g. with Maple) of explicitly presenting the free energy for
arbitrary degrees of the potentials. In the approach of \cite{KKWZ,
  MWZ1, WZ, Zab} the explicit computation in case of regions with
finite number of nonzero external moments would be prevented mainly by
the analytic computation of the logarithmic moment which, in our
approach is nothing but 
$\mu = (V_1'(Q)+V_2'(P)-PQ)_0 - t\ln \gamma^2$, where the subscript $_0$
means the constant part in $\lambda$ and $\gamma$ is the ``conformal
radius'' of the domain. This observation leads to the following 
\bc
Let $\mathcal D$ be a simply connected domain with finite number (say
$d+1$) of
nonzer external harmonic moments and let 
\bea
z(w)= \gamma w+ \sum_{j=0}^d \alpha_j w^{-j}\ ,\qquad 
S(w)=  \frac \gamma w+ \sum_{j=0}^d \ov{\alpha}_j w^{j}
\eea
be the Riemann uniformizing map and Schwarz-function
respectively. Then
\be
\frac 1 \pi\int\!\!\!\int_{\mathcal D}\ln|z|^2{\rm d}^2z =  (V'(z)+\ov{V'}(S)-zS)_0
- t\ln \gamma^2\ ,
\ee
where the subscript $_0$ means the constant part of the Laurent
polynomial in the bracket, $t=\frac 1 \pi Area(\mathcal D)$ and $V(z)=
\sum_{K=1}^{d+1} \frac {t_K}K z^K$, with $t_K$ the $K$-th external
harmonic moment.
\ec
\er
Recently \cite{KMZ} the above tau--function $\tau_\Gamma$ has been
 generalized to multiply connected domains. Also this case is
 contained in our case by imposing the same constraint on the
 potentials $V_1 = \ov {V_2} = V$. This amounts to saying that the
 curve admits an antiholomorphic involution $\varphi$ given (coarsely speaking)
 by $Q\to \ov P$ and interchanging the poles $\infty_P$ and
 $\infty_Q$. Such a curve would then become the Schottky double used
 in \cite{KMZ} and the boundary of the multiply connected domain would
 be defined implicitly by $Q(\zeta) = \ov{P (\zeta)}$ (which
 defines --in certain cases-- $g+1$ cycles on the curve $\Sigma_g$)\footnote{
An algebraic curve of genus $g$ with an antiholomorphic involution and
$g+1$ real components is called an Harnack's $M$--curve; in general a
curve with the aforementioned properties can have $k\leq g+1$ real
components, see for instance\cite{harnack}}.
The reader familiar with the results in \cite{KMZ} will have no
 difficulty in finding the necessary dictionary to translate the two
 settings, provided he/she imposes the restriction on the potentials
 in our more general case.

\subsection{Bergman kernel}
\label{bergman}
The Bergman kernel\footnote{Our use of the term ``Bergman kernel''
 is slightly unconventional, since more commonly the Bergman kernel
 is a reproducing kernel in the $L^2$ space of holomorphic
 one-forms. The kernel that we here name ``Bergman'' is sometimes
 referred to as the ``fundamental symmetric  bidifferential''. We
 borrow the (ab)use of the name ``Bergman'' from \cite{korotkin}.}
 is a classical object in complex geometry and can
be represented in terms of prime forms and Theta functions.
In fact we will not need any such sophistication because we are going
to use only its fundamental properties (that uniquely determine it).
The Bergman kernel $\Omega (\zeta,\zeta')$ (where $\zeta,\zeta'$
denote here and in the following abstract
points on the curve) is a bi-differential on $\Sigma_g\times\Sigma_g$
with the properties
\bea
&&\hbox{Symmetry: }\qquad\Omega(\zeta,\zeta')= \Omega(\zeta',\zeta)\\
&&\hbox{Normalization: }\qquad\oint_{\zeta'\in a_j}\Omega(\zeta,\zeta') = 0 \\
&&\oint_{\zeta'\in b_j} \Omega(\zeta,\zeta') = 2i\pi \omega_j(\zeta) \
= \ \hbox { the holomorphic
  normalized Abelian differential}\ .
\eea
It is holomorphic everywhere on $\Sigma_g\times\Sigma_g\setminus
\Delta$,  and it has a double pole on the diagonal $\Delta:=\{\zeta=\zeta'\}$:
namely,  if $z(\zeta)$ is any
coordinate, we have
\be
\Omega(\zeta,\zeta') \m{\simeq}_{\zeta\sim  \zeta'} \le[\frac 1{(z(\zeta)-z(\zeta'))^2}  + \frac
1 6 S_0(\zeta) + \mathcal O(z(\zeta)-z(\zeta'))\ri]{\rm d}z(\zeta){\rm d}z(\zeta')\ ,
\ee
where the very important quantity $S_0(\zeta)$ is the ``projective
connection'' (it transforms like the Schwarzian derivative under
changes of coordinates).\par
It follows also from the general theory that any normalized Abelian
differential of the third kind with simple poles at two points $z_{-}$
and $z_{+}$ with residues respectively $\pm 1$ is obtained from the
Bergman kernel as
\be
{\rm d}S_{z_+,z_-}(\zeta) = \int_{\zeta'=z_-}^{z_+}
\Omega(\zeta,\zeta')\ .
\ee
\subsubsection{Prime form}
For the sake of completeness and comparison with the results of
\cite{KMZ} we recall here that the definition of the prime form
$E(\zeta,\zeta')$.
\bd
The prime form $E(\zeta,\zeta')$ is the $(-1/2,-1/2)$ bi-differential on
$\Sigma_g\times \Sigma_g$
\bea
E(\zeta,\zeta') = \frac {\Theta\le[\alpha\atop \beta\ri]
  (\mathfrak u(\zeta)-\mathfrak u(\zeta'))}
{h_{\le[\alpha\atop \beta\ri]} (\zeta) h_{\le[\alpha\atop \beta\ri]}  (\zeta') }
\\
h_{\le[\alpha\atop \beta\ri]} (\zeta)^2 := \sum_{k=1}^{g}
\pa_{\mathfrak u_k}\Theta\le[\alpha\atop \beta\ri]\bigg|_{\mathfrak
  u=0} \omega_k(\zeta)\ ,
\eea
where $\omega_k$ are the normalized Abelian holomorphic differentials,
$\mathfrak u$ is the corresponding Abel map and $\le[\alpha\atop
  \beta\ri]$ is a half--integer odd characteristic (the prime form does
not depend on which one).
\ed
Then the relation with the Bergman kernel is the following 
\be
\Omega(\zeta,\zeta') = {\rm d}_\zeta {\rm d}_{\zeta'} \ln
E(\zeta,\zeta') = \sum_{k,j=1}^{g}\pa_{\mathfrak u_k} \pa_{\mathfrak u_j}
\Theta\le[{\alpha \atop \beta}\ri] \bigg|_{\mathfrak u(\zeta)-\mathfrak u
  (\zeta')} \omega_k(\zeta) \omega_j(\zeta')
\ee
\subsection{Second order observables}
\label{second}
The second order observables of the two-matrix model (in fact, the
multi--matrix model) have already been
investigated in the literature \cite{chaineloop, KMZ, daul} and their relation
with the Bergman kernel extensively documented. Here we just bring a
different and possibly more rigorous derivation of those identities.
The main reason why the second order observables appear already in the
literature is the expectation (and -in some cases, mostly in the
one-matrix model setting- proof)
of their ``universality'', that is their ``independence'' on the fine
details of the potentials. This paragraph will support once more  this
point of view; indeed we will see that these generalized ``specific
heats'' do not really depend on the potentials but only on the
spectral curve of the model and can be described by geometrical
objects directly linked only to the curve itself.
That is to say that these second order observables will be the same
for any pairs of potentials for which the conformal structure of the
spectral curve is isomorphic.

Of course this is no {\em proof} of universality, since there is no
(rigorous, to our knowledge) proof that the free energy as
defined in this paper is really obtained from some scaling limit of
the partition function of the matrix model. Such a proof exists in the
one--matrix model and uses rather sophisticated tools (Riemann Hilbert
problem) \cite{DKMVZ}.
A first step in this direction has already been taken by collaborators
and the author in \cite{BHE1}.\par \vskip 3pt

On the other hand we will see in Section \ref{third} that universality
does not hold for third order observables, which will not depend purely
on the conformal structure of the spectral curve but on the functions
$P$ and $Q$ explicitly.\par\vskip 5pt

Let us start with $\pa_{u_K}\pa_{u_J}\mathcal F$
\be
2i\pi \pa_{u_K}\pa_{u_J}\mathcal F = \frac 1 K 
\pa_{u_J}\oint_{\infty_Q} P Q^K{\rm d }Q =  \frac 1 K \oint_{\infty_Q}
(\pa_{u_J}P)_Q Q^K {\rm d}Q\ .
\ee
Let us focus our attention on the differential $(\pa_{u_K}P)_Q  {\rm
  d}Q$ (or $(\pa_{v_J}P)_Q {\rm d}Q$), where the subscript indicates
that the corresponding quantity is kept fixed under differentiation.
It follows from the definition of the coordinates on the moduli space
and from eqs. (\ref{cazzata}) 
that 
\bea
&&(\pa_{u_K}P)_Q  {\rm d}Q = \le(Q^{K-1} + \mathcal O(Q^{-2})\ri){\rm
  d}Q \ \ \ \hbox { near $\infty_Q$}\\
&&(\pa_{u_K}P)_Q  {\rm d}Q = -(\pa_{u_K}Q)_P{\rm d}P =  \mathcal O(P^{-2}){\rm d}P \ \ \
\hbox{near $\infty_P$}\\
&&\oint_{a_j}(\pa_{u_K}P)_Q  {\rm d}Q = \pa_{u_K}\epsilon_j = 0 \ ,
\eea
and 
\bea
&&(\pa_{v_J}P)_Q  {\rm d}Q = -(\pa_{v_J}Q)_P  {\rm d}P =  \le(-P^{J-1} + \mathcal O(P^{-2})\ri){\rm
  d}P \ \ \ \hbox { near $\infty_P$}\\
&&(\pa_{v_J}Q)_P  {\rm d}P =  \mathcal O(Q^{-2}){\rm d}Q \ \ \
\hbox{near $\infty_Q$}\\
&&\oint_{a_j}(\pa_{v_J}P)_Q  {\rm d}Q = -\pa_{v_J}\epsilon_j = 0 \ .
\eea
In these formulas we have used repeatedly (and we will use it many
times) the so--called {\em thermodynamic identity} (or {\em reciprocity}) 
\be
(\pa P)_Q{\rm d}Q = -(\pa Q)_P{\rm d}P\ ,
\ee
where $\pa$ denotes any derivative. This is immediately obtained by
differentiating $P$ as a composite function with the (local) inverse
of $Q$. \par
Note that the only singularities of these differentials  are at the two marked points (see the
discussion on the third-order observables for a proof that there are
no poles at the branch-points).
In other words $(\pa_{u_K}P)_Q  {\rm d}Q$ and $(\pa_{v_J}P)_Q  {\rm
  d}Q$ are  Abelian differentials
of the second kind (i.e. with poles but no residues), normalized
(i.e. the $a$-periods vanish) with a pole of degree $K+1$ or $J+1$
respectively  only at
$\infty_Q$ or $\infty_P$ respectively.
It is immediate that this differential is uniquely determined and can
be expressed in terms of the Bergman kernel as follows
\bea
(\pa_{u_K}P)_Q  {\rm d}Q(\zeta) =\frac 1 {2i\pi \, K }  \oint_{\infty_Q}
Q^K(\wt\zeta) \Omega(\zeta,\wt \zeta)\ ,\nonumber
\\
(\pa_{v_J}P)_Q{\rm d}Q(\zeta) = -(\pa_{v_J}Q)_P{\rm d}P = -\frac 1 {2i\pi\,
  J} \oint_{\infty_P} P^{J}(\wt \zeta) \Omega(\zeta,\wt \zeta)\ .\label{hope}
\eea
 Indeed, when $\zeta\sim \infty_Q$ we have (using $z=1/Q(\zeta)$ as local coordinate)
\bea
\frac 1 {2i\pi J }  \oint_{\infty_Q}  \!\!\!\!\!z^{-J} \Omega(\zeta,\wt
\zeta)
= \frac 1 {2i\pi J }  \oint_{\infty_Q}\!\!\!\!\!z^{-J} 
\le[\frac 1{(z- z')^2} + \mathcal O(1) \ri]{\rm
  d}z{\rm d}z' = \le(-z^{-J-1}\!\!+\!\! \mathcal O(1)\ri){\rm d}z = \le(Q^{J-1}\!\!+\!\! \mathcal O(Q^{-2})\ri){\rm d}Q   
\eea
It then follows that 
\be
\pa_{u_J}\pa_{u_K}\mathcal F = \frac 1{(2i\pi)^2 KJ}
\oint_{\infty_Q}\oint_{\infty_Q} Q^K(\zeta) Q^J(\wt\zeta) \Omega(\zeta,\wt \zeta)\ .\label{uuder}
\ee
By similar arguments one obtains also the formulas
\bea
 \pa_{u_K}\pa_{v_J}\mathcal F &=&\hspace{-8pt} \frac 1 {2i\pi K}
\oint_{\infty_Q}Q^K(\pa_{v_J}P)_Q{\rm dQ}
\m{=}^{\hbox{\tiny{thermodynamic identity}}} \\
&=&\hspace{-8pt}-\frac 1{2i\pi K}
  \oint_{\infty_Q}Q^K (\pa_{v_J}Q)_P{\rm d}P = \\
&=&\hspace{-8pt} \frac {-1}{(2i\pi)^2 
   KJ}\oint_{\infty_Q} \oint_{\infty_P}
  \Omega(\zeta,\wt \zeta)Q^K(\zeta)P^J(\wt \zeta)\label{uvder}\\
\pa_{v_J}\pa_{v_K}\mathcal F &=&\hspace{-8pt} \frac 1{(2i\pi)^2 KJ}
\oint_{\infty_P}\oint_{\infty_P} P^K(\zeta) P^J(\wt\zeta) \Omega(\zeta,\wt \zeta)\ .\label{vvder}
\eea
%The formulas for the double derivatives w.r.t. $v_J$'s are analogous
%to eq. (\ref{uuder}) where the r\^ole of $P$ and $Q$ and their poles
%is interchanged.\\
We now compute the second derivatives w.r.t. the filling fractions; to
that extent we notice that $(\pa_{\epsilon_j}P)_Q{\rm d}Q$ is
holomorphic everywhere because
\be
(\pa_{\epsilon_j}P)_Q{\rm d}Q = \le\{\begin{array}{ll}
\ds \mathcal O(Q^{-2}){\rm d}Q  &\hbox{near $\infty_Q$}\cr
\ds = -(\pa_{\epsilon_j}Q)_P{\rm d}P = \mathcal O(P^{-2}){\rm d}P &
\hbox{near $\infty_P$}\ .
\end{array}
\ri.
\ee
Thus it is regular at the marked points and has no other singularities
(the poles of the variation at $Q$ fixed of $P$ cancel with the zeroes
of ${\rm d}Q$; see discussion in section \ref{third}).
Moreover it satisfies
\be
  2i\pi \delta_{jk} = 2i\pi \pa_{\epsilon_j}\epsilon_k =
  \pa_{\epsilon_j}\oint_{a_k} P{\rm d}Q=
  \oint_{a_k}(\pa_{\epsilon_j}P)_Q{\rm d}Q\ .
\ee
Therefore $(\pa_{\epsilon_j}P)_Q{\rm d}Q= {2i\pi}\omega_j$
where $\{\omega_j\}_{j=1\dots g}$ is the basis of normalized Abelian holomorphic
differentials. As a consequence we obtain
\be
\pa_{\epsilon_j}\pa_{\epsilon_k} \mathcal F = 
\oint_{b_k}(\pa_{\epsilon_j} P)_Q{\rm d}Q = {2i\pi}\oint_{b_k}\omega_j = 
     \oint_{b_k}\oint_{b_j} \Omega = 2i\pi \, \tau_{jk} \ ,
\ee
where ${\bf \tau }= [\tau_{ij}]_{i,j=1\dots g}$ is the period matrix
for the holomorphic curve.

The other mixed derivatives are easily computed along the same lines
as above to be 
\bea
&&\pa_{u_K}\pa_{\epsilon_j}\mathcal F = \frac 1 {2i\pi K}
\oint_{\zeta\in b_j} \oint_{\infty_Q}
\Omega(\zeta,\zeta')Q^K(\zeta')\\
&& \pa_t\pa_{u_K}\mathcal F=\frac {-1}{2i\pi K} \oint_{\infty_Q} Q^K{\rm d}S\\
&& \pa_t\pa_{v_J}\mathcal F = \frac 1{2i\pi J}\oint_{\infty_P}P^J{\rm
  d}S\\
&& \pa_t\pa_{\epsilon_j}\mathcal F = \oint_{b_j}{\rm d}S =
\mathfrak u_j(\infty_Q)-\mathfrak u_j(\infty_P)\\
&& \pa_t^2\mathcal F = \ln(\gamma\widetilde \gamma) ,\\ 
&& \ln(\gamma):=-
\res{\infty_Q} \ln\le(\frac Q\lambda\ri){\rm d}S\ ,\ \ \ln(\wt\gamma):=
\res{\infty_P} \ln\le(P\lambda \ri){\rm d}S\ .
\eea
(Note that the product $\gamma\wt\gamma$ does not depend on the
arbitrary multiplicative constant entering the definition of the
multivalued function $\lambda$).
The first two relations are obvious recalling that \cite{F1} 
\be
{\rm d} S = -(\pa_t P)_Q{\rm d}Q = (\pa_t Q)_P{\rm d}P\ .
\ee 
The third relation is a well known result for the third--kind Abelian differential.\\
We shall prove the last relation: we start from the first derivative
of $\mathcal F$ ($\mu$, also called chemical potential) and its expression
given in our previous paper \cite{F1}
\bea
\hspace{-20pt}2i\pi\mu = 2i\pi\pa_t\mathcal F
\hspace{-9pt} &=&\hspace{-10pt}
\oint_{\infty_Q}\hspace{-8pt}\le(V_1(Q)-t\ln\le(Q/\lambda\ri)\ri){\rm d}S -
\oint_{\infty_P}\hspace{-8pt}\le( V_2(P)-t\ln(P\lambda)\ri){\rm d} S -
\oint_{\infty_Q}\hspace{-8pt}PQ{\rm d}S +2i\pi \sum_{j=1}^g \epsilon_j\oint_{b_j}\hspace{-5pt}
     {\rm d}S =\\ &=&\hspace{-10pt}\oint_{\infty_Q}\hspace{-8pt}\ln\le(\frac Q \lambda\ri)P{\rm d}Q +
     \oint_{\infty_P}\hspace{-8pt}\ln\le(P\lambda\ri) Q{\rm d}P -\oint_{\infty_Q}\hspace{-8pt}PQ{\rm d}S +2i\pi \sum_{j=1}^g \epsilon_j\oint_{b_j}
     \hspace{-4pt}{\rm d}S
\eea

%When computing
%the derivative of $\mu$ we can discard the terms $\oint_{b_j}{\rm d}S=
%\oint_{b_j} \frac {{\rm d}\lambda}\lambda$ because the variation can
%be done at $\lambda$ fixed. Therefore we are left with
%
%
%
We can therefore compute
\bea
2i\pi\pa_t\mu &=& -\oint_{\infty_Q} \frac 1
\lambda\le(\pa_t\lambda\ri)_QP{\rm d}Q + \overbrace{\oint_{\infty_Q}\ln\le(\frac
Q\lambda\ri) \overbrace{(\pa_tP)_Q{\rm d}Q}^{=-{\rm d}S}}^{:=2i\pi\ln \gamma}+\\
&&+ \oint_{\infty_P}\frac 1 \lambda \le(\pa_t\lambda \ri)_P
Q{\rm d}P + \overbrace{\oint_{\infty_P}\ln\le(P\lambda\ri)\overbrace{ (\pa_tQ)_P
  {\rm d}P}^{={\rm d}S}}^{=2i\pi\ln\widetilde\gamma}
+\label{rds}\\
&&-\oint_{\infty_Q}P \overbrace{(\pa_t Q)_\lambda\frac {{\rm
      d}\lambda}\lambda}^{=-\frac{(\pa_t\lambda)_Q}\lambda{\rm d}Q}
-\oint_{\infty_Q} Q
\overbrace{(\pa_t
P)_\lambda\frac {{\rm d}\lambda}{\lambda}}^{=-\frac{(\pa_t\lambda)_P}\lambda
  {\rm d}P} + 2i\pi\sum_{j=1}^{g} \epsilon_j\oint_{b_j} (\pa_t{\rm d}S)_P=\\
&=& 2i\pi\ln(\gamma\widetilde\gamma)\ ,
\eea 
where we have used Riemann bilinear identity to move the residue of
the following term on line (\ref{rds})
\be
 \oint_{\infty_P}\frac 1 \lambda \le(\pa_t\lambda \ri)_P
Q{\rm d}P = - \oint_{\infty_P}\frac 1 \lambda \le(\pa_t\lambda \ri)_P
Q{\rm d}P  + \sum_{j=1}^g\overbrace{ \oint_{a_j} Q{\rm d}P
}^{=-2i\pi\epsilon_j}\le[ \oint_{b_j}(\pa_t{\rm d}S)_P - \oint_{b_j}
  Q{\rm d}P \underbrace{\oint_{a_j}(\pa_t {\rm d}S)_P}_{=\pa_t
    \oint_{a_j} \!\!\!{\rm d}S =0}\ri]\ .
\ee
This concludes our analysis of the second order observable of the
model. 
We repeat here for the sake of completeness  that we are aware of a
prior derivation in \cite{chaineloop} and \cite{WZ2} for the case of
conformal maps and genus $0$ (one--cut case \cite{daul}). 

\subsection{Extension to formal power series for the potentials}
\label{formal}
The present derivation for the second derivatives does not rely on
formal manipulations involving higher $u_K$ or $v_J$. However it gives
a deeper insight to use such a formalism.
To that end we will now think of the two potentials $V_1$ and $V_2$ as
infinite power series, without concern about their convergence.
This is often done in the physical literature
 \cite{Bertrand_unpublished, MWZ1} and yields the correct results
in a faster and possibly more elegant way.
To this end we  define the {\em puncture} operators
\be
\frac{\delta}{\delta V_1(q)} := \sum_{K=1}^{\infty} q^{-K-1}K\pa_{u_K}
\ ;\qquad \frac{\delta}{\delta V_2(p)} := \sum_{J=1}^{\infty} p^{-J-1}J\pa_{u_J}
\ee
Using this formalism we realize that our free energy $\mathcal F_g$ is nothing but the
{\em generating function} of the Bergman kernel on the selected curve
\be
\frac{\delta^2 \mathcal F}{\delta V_1(q)\delta V_1(q')}
     = -\frac 1
     {4\pi^2}\oint_{\infty_Q}\oint_{\infty_Q}\frac
     {\Omega(\zeta,\zeta')}{(q-Q(\zeta))(q'-Q(\zeta'))} = \frac
     1{(q-q')^2} + 
     \frac{\Omega(\zeta(q),\zeta(q'))}{{\rm d}Q(\zeta(q)){\rm d}Q(\zeta(q'))}\ ,\label{fancy}
\ee
where the integrals are to be read in a formal sense of inverse power
series in $q,q'$, coefficient by coefficient; indeed each coefficient
is precisely given by formula (\ref{uuder}).
In eq. (\ref{fancy}) the notation $\zeta(q)$ means the point on the
curve that projects on the {\em physical} sheet of the projection
$Q$. The {\em physical } sheet is the sheet that has no cuts extending
to infinity. 
On the physical sheet the formula is not formal; on the other sheets
it should be properly understood as analytic continuation.

Similarly we have the other formulas where the $P$-projection is involved
\bea
&& \frac{\delta^2 \mathcal F}{\delta V_2(p)\delta V_2(p')}
     {\rm d}p\,{\rm d}p'= -\frac 1
     {4\pi^2}{\rm d}p\,{\rm d}p'\oint_{\infty_P}\oint_{\infty_P}\frac
     {\Omega(\zeta,\zeta')}{(p-P(\zeta))(p'-P(\zeta'))} =\frac
     1{(p-p')^2} +
    \frac{\Omega(\wt\zeta(p),\wt\zeta(p'))}{{\rm
	d}P(\wt\zeta(p)){\rm d}P(\wt\zeta(p'))}  ,\\
&&\frac{\delta^2 \mathcal F}{\delta V_1(q)\delta V_2(p')}
     {\rm d}q\,{\rm d}p'= -\frac 1
     {4\pi^2}{\rm d}q {\rm d}p' \oint_{\infty_Q}\oint_{\infty_P}\frac
     {\Omega(\zeta,\zeta')}{(q-Q(\zeta))(p'-P(\zeta'))} =
     \frac{\Omega(\zeta(q),\zeta(p'))}{{\rm d}Q(\zeta(q)){\rm
	 d}P(\wt\zeta(p'))}\ .
\eea
To be more precise, these formulas are exact (modulo the issue of
convergence of the potentials) only in a neighborhood of the
respective punctures where $Q$ and $P$ provide univalued
coordinates.\par
It is interesting to recast the problem upside-down: let it be given a
smooth curve of genus g $\Sigma_g$ with Bergman kernel $\Omega$ (which implies a fixation
of symplectic basis in the homology of the curve) and two marked
points $\infty_{Q,P}$ such that:\\
there exist two functions $Q,P$ with simple poles at the respective
marked points;\\
the only other singularity for the functions above are essential
singularities at the other (respective) marked point.\par
In this setting we can define ``coordinates'' $u_K$, $v_J$ by means of
formulas (\ref{moduli}) and then the expansion of the Bergman kernel
around the punctures in the coordinates provided by the functions
$P,Q$ are the second derivatives of a ``tau'' function.
\subsubsection{Relation with conformal maps}
The reader trying to bridge the present setting with \cite{KMZ} should be warned that
the puncture operator here is {\em not} the ``infinitesimal
deformation'' operator ibidem. Their operator would be written in our
notation
\be
\frac {\delta}{\delta V'_1(x)} = \sum_{K=1}^\infty x^{-K}\pa_{u_K} \ ,\qquad \frac {\delta}{\delta
  V_2'(y)} = \sum_{J=1}^\infty y^{-J}\pa_{v_J} \ ,
\ee
with the relation (e.g.  for the $x-x$ two-puncture case)
\be
\frac {{\rm d}}{{\rm d}x_1} \frac {{\rm d}}{{\rm d}x_2} \frac {\delta}{\delta V'_1(x_1)}\frac {\delta}{\delta V'_1(x_2)}
\mathcal F_g= \frac {\delta}{\delta V_1(x_1)}\frac {\delta}{\delta V_1(x_2)}
\mathcal F_g  
\ee
Therefore their double-deformation is the antiderivative of our double
puncture. This is why in \cite{KMZ} they have the logarithm of the
prime form\footnote{Here we are speaking rather loosely; we should also fix a coordinate and consider the
  prime form as a bi-function by trivializing the spinor bundle.}
whereas we have the Bergman kernel. Since in this paper we will not
investigate Hirota equations (i.e. Fay's identities) which involve
the prime form more naturally than the Bergman kernel, we will not
pursue this observation any further.
\section{Third derivatives: Beltrami differentials}
\label{third}
In this section we compute the third derivatives in terms of the canonical
structures and data of our moduli space.
We must note here that in the case of the tau-function for conformal
maps of connected domains (i.e. genus $0$ case)  constructed by Marshakov, Krichever, Wiegmann, Zabrodin et. al. these
formulas are known \cite{MWZ1} (see appendix \ref{strnzn}). In order to obtain their ``residue
formulas'' from the formulas we are going to write down, we
should restrict ourselves to the genus zero case and impose that $P(\lambda) =
\ov Q(1/\ov \lambda)$, $\lambda$ being the uniformizing parameter of
the rational curve with $\lambda(\infty_Q)=\infty$ and
$\lambda(\infty_P)=0$ (see appendix \ref{strnzn}). However  our formulas cover a more general
case that would includes the situation of conformal maps of
multiply--connected domains \cite{KMZ}, were we to impose some
restriction of reality on $P$ and $Q$ (as explained in section \ref{setting})

%In what follows we will compute variations at $Q$ fixed but everything
%could be done keeping $P$ fixed. The resulting formulas should be
%independent of the choice.\par
When computing a variation in the moduli of our problem $\{u_K, v_J,
t,\epsilon_j\}$,  we introduce a deformation of the conformal
structure of the curve.
We will see that this deformation is equivalent 
to introducing simple poles
at the branch-points of one or the other of the  two projections $Q:\Sigma_g\to \C P^1$ and
$P:\Sigma_g \to\C P^1$, depending on which among $Q$ and $P$ we keep
fixed under the variation.
Let us denote by $q_\mu, \mu=1,.\dots,d_2+1+2g$ and $p_\nu,\
\nu=1,.\dots, d_1+1+2g$  such points and
then by $Q_\mu, \ P_\nu$ the corresponding critical values;
here we are implicitly making an assumption of genericity of the
potentials, in that we will consider these branch-points as
simple. Were we to consider more degenerate cases  ({\bf critical
  potentials}) we would have to
modify some of the formulas below (mainly the Beltrami
differentials). Note that critical potentials (i.e. potentials which
are fine-tuned so as to have degenerate singularities of the maps $P$
and $Q$) are very important for applications to conformal field
theories in that they provide instances of minimal conformal models.\par
Coming back to our generic situation,  
let us consider the case of the $Q$-projection, the other case being
totally similar. 
The projection $Q:\Sigma_g\to \C P^1$ determines the conformal
structure of the curve and hence it suffices to consider the variation
of the structure arising from the deformation of this projection.
Consider first the branch-points of the $Q$-projection and introduce
the local parameter in a neighborhood of the critical point $q_\mu$
\be
\zeta_\mu = \sqrt{Q-Q_\mu}\ .
\ee
As we know, the critical values of both projections $Q$ and $P$ are {\em not} independent coordinates
in our moduli space;  they  vary when performing a
variation of our moduli.
Let us call $\pa$ any infinitesimal variation of our moduli. 
Under such a variation $Q_\mu$ also undergoes some variation $\pa
Q_\mu$. Suppose now we have a germ of function $F$ in a neighborhood
of $q_\mu$ (which may also depend explicitly on the moduli) and we make the variation $\pa$ at $Q$-value fixed. Then
\bea
(\pa F(\zeta_\mu))_Q =  -\frac {\pa Q_\mu}{2\zeta_{\mu}} \frac{
  {\rm d}F(\zeta_\mu)}{{\rm d}\zeta_\mu} + \pa_\pa F(\zeta_{\mu})
\label{brunch}\ ,
\eea
where $\pa_\pa F$ denotes the variation of $F$ coming from the
explicit dependence on the moduli.
We see immediately from (\ref{brunch}) that, generically, the variation has a simple pole at each
branch-point of the $Q$-projection. Ditto for the
branch-points of $P$ when computing a variation at $P$--value fixed.\par\vskip 6pt
A variation of the moduli induces a variation in the conformal
structure of the underlying real surface $\Sigma_g$; these
infinitesimal variations are described by the so-called ``Beltrami
differentials''. 
They consist of differentials of the form 
\be
\mu(z,\ov z) {\rm d}\ov z/{\rm d}z \ \in L^{\infty}(\Sigma_g)\ ,
\ee
and enter into Rauch's formula \cite{rauch} for the variation of the
Bergman kernel
\be
\delta_\mu \Omega(\zeta,\zeta') = \int\int_{\Sigma_g}
\mu(\zeta'')\Omega(\zeta'',\zeta)\Omega(\zeta'',\zeta')\ .\label{Rauch}
\ee 
It is shown in many places \cite{Fay, rauch, korotkin,korotkin2} that varying
the image of a branch-point $b$  of a covering $\varphi:\Sigma_g\to \C P^1$ corresponds to the
following Beltrami differential (in terms of a local coordinate $z$
centered at $b$)
\be
\mu_{Sc} = -\frac 1 {2\epsilon^2} \1_{ \epsilon }(|z|)\frac {{\rm d}
  \ov z}{{\rm d}z}\ ,
\ee
where  $\1_{ \epsilon }$ is the characteristic function of the
$\epsilon$ disc and $\epsilon$ is a parameter small enough so that no
other branch-points fall within the disc. This is called ``Schiffer'' variation and
corresponds to varying the critical value of a simple branch-point; for higher
order branch-point see \cite{korotkin}.
 Plugging this into (\ref{Rauch}) and using Green's theorem followed
 by Cauchy's residues theorem one
 obtains simply a residue 
\be
\delta_{\mu_{Sc}} \Omega(\zeta,\zeta') = \res{b}
\frac{\Omega(\zeta,\zeta'')\Omega(\zeta',\zeta'')}{{\rm d}
  \varphi(\zeta'')}\ .
\ee 

Our setting is slightly different, in that we have {\em two}
projections $P$ and $Q$ and the respective branching values are not
independent moduli but vary together with  the
$\{u_K, v_J,\epsilon_j,t\}$ moduli of our problem. 
Therefore all we need in order to find the Beltrami differentials corresponding
to these variations is to be able to find the coefficients $\pa
Q_\mu$ and $\pa P_\nu$  that appear in eq. (\ref{brunch}).
To this end let us now consider the case of a differential of the form $
F {\rm d}Q$.
It follows from the above discussion (and we have already used this fact
implicitly to compute the first variations of $P{\rm d}Q$) that its variation at $Q$ fixed will be
holomorphic at the branch-points because the differential ${\rm d}Q$
has simple zeroes there which cancel the poles of the variation of $F$.

Suppose now that for some reason we have a way of identifying
independently what differential is $\omega := (\pa F)_Q{\rm d}Q$
Then we could compute the desired
coefficients $\pa Q_\mu$ as in the following formula
\bea
&& \frac{\omega}{{\rm d}\zeta_\mu} (q_\mu)  = -\pa Q_\mu \frac {{\rm
    d}F}{{\rm d}\zeta_\mu} (q_\mu)\ \hbox{ that is...}\\
&& \pa Q_\mu = -\frac{\omega}{{\rm d}F} (q_\mu)
\eea
We now apply the above to the case $F=P$ and the differential $P{\rm
  d}Q$; in this case we have already discovered that 
\begin{enumerate}
\item the derivative of $P{\rm d}Q$ w.r.t. $\epsilon_j$
  is the
  holomorphic normalized Abelian differential    $(\pa_{\epsilon_j}P)_Q{\rm d}Q = \omega_j$;
\item the derivatives of $P{\rm d}Q$ w.r.t. $u_K$ or $v_J$ are given
  by eq. (\ref{hope});
\item the derivative of $P{\rm d}Q$ w.r.t. $t$ is  the normalized
  Abelian integral $(\pa_t P)_Q{\rm d}Q ={\rm d}S$  of the third kind with poles at the marked points.
\end{enumerate}
Therefore the parameter $\pa Q_\mu$ and $\pa P_\nu$ are given by
the following formulas
\bea
\pa_{\epsilon_j} Q_\mu &=&-\le. \frac {\omega_j}{{\rm d}P} \ri|_{q_\mu} =- \frac
1 {2i\pi} \frac {\ds \oint_{\wt \zeta \in  b_j} \Omega(\zeta,\wt \zeta)}{{\rm d}P(\zeta)}\bigg|_{\zeta=q_\mu} \\
\pa_{u_K} Q_\mu &=& -\frac  1{2i\pi K} \frac {\ds \oint_{\infty_Q}
Q(\wt \zeta)^K\Omega(\zeta,\wt \zeta)} {{\rm d}P(\zeta)}\bigg|_{\zeta=q_\mu}\\
\pa_{v_J} Q_\mu &=& \frac 1{2i\pi J}\frac{\ds \oint_{\infty_P}
  P(\wt \zeta)^J\Omega
(\zeta,\wt \zeta)}{{\rm d}P(\zeta)} \bigg|_{\zeta=q_\mu}\\
\pa_t Q_\mu &=&-\frac {\ds \int_{\infty_Q}^{\infty_P} \Omega(\zeta,\wt
  \zeta)}{{\rm d}P(\zeta)}\bigg|_{\zeta=q_\mu}\ .
\eea
Repeating all of the above but interchanging the r\^ole of $P$ and $Q$
we find (note the opposite signs, coming from the thermodynamic identity)
\bea
\pa_{\epsilon_j} P_\nu &=& \frac {\omega_j}{{\rm d}Q} \bigg|_{p_\nu} = \frac
1 {2i\pi} \frac {\ds \oint_{\wt \zeta \in  b_j} \Omega(\zeta,\wt \zeta)}{{\rm d}Q(\zeta)}\bigg|_{\zeta=p_\nu} \\
\pa_{u_K} P_\nu &=& \frac  1{2i\pi K} \frac {\ds \oint_{\infty_P}
Q(\wt \zeta)^K\Omega(\zeta,\wt \zeta)} {{\rm d}Q(\zeta)}\bigg|_{\zeta=p_\nu}\\
\pa_{v_J} P_\nu &=& -\frac 1{2i\pi J}\frac{\ds \oint_{\infty_P}
  P(\wt \zeta)^J\Omega
(\zeta,\wt \zeta)}{{\rm d}Q(\zeta)} \bigg|_{\zeta=p_\nu}\\
\pa_t P_\nu &=&  \frac {\ds \int_{\infty_Q}^{\infty_P} \Omega(\zeta,\wt
  \zeta)}{{\rm d}Q(\zeta)}\bigg|_{\zeta=p_\nu}\ .
\eea
In general we can write the formula ($\pa$ denoting an arbitrary
variation)
\bea
\pa Q_\mu = -\frac {(\pa P)_Q {\rm d}Q}{{\rm d}P}\bigg|_{q_\mu} = (\pa
Q)_P(q_\mu)\\
\pa P_\nu = -\frac {(\pa Q)_P {\rm d}P}{{\rm d}Q}\bigg|_{p_\nu} = (\pa
P)_Q(p_\nu)\ .
\eea
We are now in a position of expressing the Beltrami differentials corresponding to a
variation in our moduli as in 
\be
\mu_{(\pa)} = \sum_{\mu = 1}^{d_2+1 + 2g} \pa Q_\mu
\mu_{Sc}^{(q_\mu)} \equiv  \sum_{\nu = 1}^{d_1+1 + 2g} \pa P_\nu
\mu_{Sc}^{(p_\nu)}\ ,
\ee
where the notation $\mu_{Sc}^{(\zeta)}$ denotes the Schiffer variation
centered at the point $\zeta$.
The equivalence sign $\equiv$ means that the two expressions determine
the same variation of conformal structure. Indeed we should remind that any two
Beltrami differentials $\mu$ and $\wt\mu$ determine the same variation
of the conformal structure if for any quadratic holomorphic
differential $\Phi$ we have 
\be
\int\!\!\!\!\int_{\Sigma_g} \mu \Phi = \int\!\!\!\!\int_{\Sigma_g}\wt\mu\,\Phi\ .
\ee
Indeed, for $\Phi\in H^0(\mathcal K^2)$ a holomorphic quadratic differential we have
\bea
&& \sum_{\mu = 1}^{d_2+1 + 2g} \pa Q_\mu \int\!\!\int_{\Sigma_g}
\mu_{Sc}^{(q_\mu)} \,\Phi = 
\sum_{\mu=1}^{d_2+1+g}
\pa Q_\mu \res{\zeta=q_\mu} \frac \Phi {{\rm d}Q} = - \sum_{\mu=1}^{d_2+1+g}
 \res{\zeta=q_\mu} \frac {(\pa P)_Q{\rm d}Q \,\Phi} { {\rm d}Q {\rm
       d}P} \m{=}^{(\star)} \\
&& = \sum_{\nu=1}^{d_1+1+g}
 \res{\zeta=p_\nu} \frac {(\pa P)_Q{\rm d}Q \,\Phi}{ {\rm d}Q {\rm
       d}P} = \sum_{\nu=1}^{d_1+1+g}
\pa P_\nu \res{\zeta=p_\nu} \frac \Phi {{\rm d}P} = 
 \sum_{\nu = 1}^{d_1+1 + 2g} \pa P_\nu\int\!\!\!\!\int_{\Sigma_g}
 \mu_{Sc}^{(p_\nu)} \Phi\ ,  
\eea
thus proving the equivalence of the two Beltrami differentials.
The equality marked $(\star)$ follows from the fact that the
  differential $ \frac {(\pa P)_Q{\rm d}Q \,\Phi} { {\rm d}Q {\rm
       d}P}$ has poles only at the branch-points of $P$ and $Q$, and
  the sum of all residues must vanish\footnote{The differentials $(\pa
    P)_Q{\rm d}Q$ have poles at most of degree $d_2+2$ at $\infty_Q$
    or $d_1+2$ at $\infty_P$ but those are canceled by the poles of
    ${\rm d}P{\rm d}Q$ at those points, of degree $d_2+3$ and $d_1+3$ respectively.}.
%Note that the two Beltrami differentials are equivalent only for
%  variations of holomorphic structure, that is the equivalence would
%fail were we to consider meromorphic quadratic differentials.

From Rauch variational formula (\ref{Rauch}) (see also
\cite{korotkin}) follows then that  
the variation of the Bergman kernel under such a Beltrami
differential is
\bea
&&(\pa \Omega(\zeta,\zeta'))_Q = \sum_{\mu=1}^{d_2+1+2g} \res{\wt \zeta=q_\mu}
\pa Q_\mu \frac
       {\Omega(\zeta,\wt \zeta)\Omega (\zeta',\wt \zeta)}{{\rm d}Q(\wt
	 \zeta)}   = \sum_{\mu=1}^{d_2+1+2g} \res{\wt \zeta=q_\mu} \frac
       {(\pa P)_Q{\rm d}Q(\wt\zeta)\Omega(\zeta,\wt \zeta)\Omega
	 (\zeta',\wt \zeta)}{{\rm d}P(\wt\zeta){\rm d}Q(\wt
	 \zeta)}\\
&&(\pa \Omega(\zeta,\zeta'))_P = 
 \sum_{\nu=1}^{d_1+1+2g} \res{\wt \zeta=p_\nu}
\pa P_\nu \frac
       {\Omega(\zeta,\wt \zeta)\Omega (\zeta',\wt \zeta)}{{\rm d}P(\wt
	 \zeta)} =  
 \sum_{\nu=1}^{d_1+1+2g} \res{\wt \zeta=p_\nu}
\frac
       {(\pa Q)_P{\rm d}P(\wt\zeta)\Omega(\zeta,\wt \zeta)\Omega (\zeta',\wt \zeta)}{{\rm d}P(\wt
	 \zeta){\rm d}Q(\wt\zeta)}\ .
\eea

Using the expressions above for the coefficients $\pa Q_\mu$ and
$\pa P_\nu$ we
obtain then
\bea
&&( \pa_{\epsilon_j}\Omega(\zeta,\zeta'))_Q =- \frac
1{2i\pi} \sum_{\mu=1}^{d_2+1+2g}
\res{\wt \zeta = q_\mu} \frac{ \ds \oint_{\zeta''\in b_j}\Omega(\zeta,\wt
  \zeta)\Omega(\wt \zeta,\zeta'')\Omega(\wt \zeta,\zeta')}
{{\rm d}P(\wt \zeta){\rm d}Q(\wt  \zeta)}\\
&& (\pa_{u_K}\Omega (\zeta,\zeta'))_Q =  -\frac 1{2i\pi K} \sum_{\mu=1}^{d_2+1+2g}
\res{\wt \zeta =q_\mu} \frac{ \ds \oint_{\zeta''\sim \infty_Q}Q(\zeta'')^K\Omega(\zeta,\wt
  \zeta)\Omega(\wt \zeta,\zeta'')\Omega(\wt \zeta,\zeta')}{{\rm d}P(\wt \zeta){\rm d}Q(\wt
  \zeta)}\\
&&( \pa_{v_J}\Omega (\zeta,\zeta'))_Q =  \frac 1{2i\pi J} \sum_{\mu=1}^{d_2+1+2g}
\res{\wt \zeta = q_\mu} \frac{ \ds \oint_{\zeta''\sim \infty_P}P(\zeta'')^J\Omega(\zeta,\wt
  \zeta)\Omega(\wt \zeta,\zeta'')\Omega(\wt \zeta,\zeta')}{{\rm d}P(\wt \zeta){\rm d}Q(\wt
  \zeta)}\\
&& (\pa_{t}\Omega (\zeta,\zeta'))_Q =  -\sum_{\mu=1}^{d_2+1+2g}
\res{\wt \zeta = q_\mu} \frac{ \ds \int_{\zeta''= \infty_Q}^{\infty_P}\Omega(\zeta,\wt
  \zeta)\Omega(\wt \zeta,\zeta'')\Omega(\wt \zeta,\zeta')}{{\rm d}P(\wt \zeta){\rm d}Q(\wt
  \zeta)} \ .
\eea
The variations at $P$ fixed would be given by the same formulas as above
but replacing the sum over the critical point of $Q$ by the sum over
the critical points of $P$ and changing the overall sign.\par
The reason why we distinguish between the variations at $Q$ or $P$
fixed will be clearer in a moment.\\
Indeed let us compute now the third derivatives of $\mathcal F_g$.
If for example we want to compute $\pa_{u_J}\pa_{u_L}
\pa_{v_K}\mathcal F_g$ the simplest way is to leave the variation of
$v_K$ last so that we will be varying at $Q$ fixed;
\bea
&& \pa_{u_J}\pa_{u_L}
\pa_{v_K}\mathcal F_g = \pa_{v_K}\pa_{u_J}\pa_{u_L} \mathcal F_g=\\
&&= \frac 1 {2i\pi} \pa_{v_K}
\oint_{\infty_Q}\!\!\!\! \frac{Q^K}K (\pa_{u_L} P)_Q{\rm d}Q =
\frac 1{(2i\pi)^2 JL}
\pa_{v_K} \!\!\!\oint_{\infty_Q}\!\! \oint_{\infty_Q}\!\!\!\! Q(\zeta)^J Q(\zeta')^L
\Omega(\zeta,\zeta') =\\
&&=  -\frac 1{4\pi^2 JL}
\oint_{\infty_Q} \oint_{\infty_Q} Q(\zeta)^J Q(\zeta')^L
(\pa_{v_K} \Omega(\zeta,\zeta'))_Q =\\
&&=  -\frac 1{8i\pi^3 JLK}\sum_{\mu=1}^{d_2+1+2g}
\res{\wt\zeta=q_\mu}\oint_{\infty_Q} \oint_{\infty_Q} \oint_{\infty_P} Q(\zeta)^J
Q(\zeta')^L P(\zeta'')^K 
 \frac{ \Omega(\zeta,\wt\zeta)
\Omega(\zeta',\wt\zeta)\Omega(\zeta'',\wt\zeta)}{{\rm
     d}P(\wt\zeta){\rm d}Q(\wt\zeta)} = \\
&&= \sum_{\mu=1}^{d_2+1+2g} \res{\wt \zeta=q_\mu} \frac
{(\pa_{v_K}P)_Q{\rm d}Q (\pa_{u_J}P)_Q{\rm d}Q (\pa_{u_L}P)_Q{\rm d}Q
}{{\rm d}Q {\rm d}P}\ .
\eea
Performing the variation at $P$ fixed, since $P$ is not a local
 coordinate near the marked point $\infty_Q$, would introduce
 unnecessary complication to the formula (basically adding an extra
 residue to it).
Thus we need
to pay the necessary care in choosing which among $P$ and $Q$ needs to
be kept fixed in order to have the simplest formulas.\par
 With this {\em caveat} in mind can thus now compute the third derivatives of the free energy. The resulting formulas
should be compared with the residue formulas in \cite{MWZ1, WZ2}
(which would correspond to genus $g=0$ here)
The formulas are simplified if we introduce the quadri-differential 
\be
\Omega^{(3,1)}(\zeta_1,\zeta_2,\zeta_3;\zeta)  = \frac {\Omega(\zeta_1, \zeta)\Omega
  (\zeta_2,\zeta) \Omega(\zeta_3,\zeta)}{{\rm d}P(\zeta){\rm
    d}Q(\zeta)}\ ,
\ee
and the notations
\bea
\mathcal U_K(\zeta):= (\pa_{u_K}P)_Q{\rm d}Q(\zeta) = \frac
1{2i\pi K}\oint_{\infty_Q} Q(\zeta')^K \Omega(\zeta,\zeta')\ ,\\
\mathcal V_J(\zeta):= (\pa_{v_J}P)_Q{\rm d}Q(\zeta)= -\frac
1{2i\pi J}\oint_{\infty_P} P(\zeta')^J \Omega(\zeta,\zeta')\ .
\eea
We also recall that 
\be
\omega_j = (\pa_{\epsilon_j}P)_Q{\rm d}Q\ ,\qquad {\rm d}S = -(\pa_t
P)_Q{\rm d}Q\ ,
\ee
are the normalized Abelian differentials of the first and third kind respectively.
With these definitions and reminders we have
\bea
\pa_{\epsilon_i}\pa_{\epsilon_j}\pa_{\epsilon_k} \mathcal F_{g}  &=& -
\sum_{\mu=1}^{d_2+1+2g} \res{\zeta=q_\mu} 
\frac
    {\omega_i(\zeta)\omega_j(\zeta)\omega_k(\zeta)}{{\rm d}P{\rm d}Q}
    = \\
& =&
-\frac 1{(2i\pi)^3} \sum_{\mu=1}^{d_2+1+2g}\hspace{-6pt}\res{ \zeta = q_\mu} \oint_{\zeta_1\in b_i}\oint_{\zeta_2\in
  b_j}\oint_{\zeta_3 \in b_k}\hspace{-5pt}
\Omega^{(3,1)}(\zeta_1,\zeta_2,\zeta_3;\zeta)= \\
& =&
\sum_{\nu=1}^{d_1+1+2g} \res{\zeta=p_\nu} \frac
    {\omega_i(\zeta)\omega_j(\zeta)\omega_k(\zeta)}{{\rm d}P{\rm
	d}Q}=2i\pi \pa_{\epsilon_i}\tau_{jk}\ .
\eea
\bea
\pa_{\epsilon_i}\pa_{\epsilon_j}\pa_{t} \mathcal F_{g} &=&  - \frac
1{(2i\pi)^2}  \sum_{\mu=1}^{d_2+1+2g}\hspace{-6pt}\res{ \zeta = q_\mu} \oint_{\zeta_1\in b_i}\oint_{\zeta_2\in
  b_j}\int_{\zeta_3=\infty_Q}^{\infty_P}
\Omega^{(3,1)}(\zeta_1,\zeta_2,\zeta_3;\zeta) = \\
 &=& \sum_{\mu=1}^{d_2+1+2g}\hspace{-6pt}\res{ \zeta =
  q_\mu}\frac{\omega_i(\zeta)\omega_j(\zeta){\rm d}S(\zeta)}{{\rm
    d}P(\zeta){\rm d}Q(\zeta)} =\sum_{\nu=1}^{d_1+1+2g}\hspace{-6pt}\res{ \zeta =
  p_\nu}\frac{\omega_i(\zeta)\omega_j(\zeta){\rm d}S(\zeta)}{{\rm
    d}P(\zeta){\rm d}Q(\zeta)} 
\eea
\bea
\pa_{\epsilon_i}\pa_{t}^2 \mathcal F_{g} &=& -\frac
1{2i\pi}  \sum_{\mu=1}^{d_2+1+2g}\hspace{-6pt}\res{ \zeta = q_\mu}
\oint_{\zeta_1\in b_i}\int_{\zeta_2=\infty_Q}^{\infty_P}
 \int_{\zeta_3=\infty_Q}^{\infty_P}
\Omega^{(3,1)}(\zeta_1,\zeta_2,\zeta_3;\zeta) = \\
&=& - \sum_{\mu=1}^{d_2+1+2g}\hspace{-6pt}\res{ \zeta =
  q_\mu}\frac{\omega_i(\zeta){\rm d}S(\zeta)^2}{{\rm
    d}P(\zeta){\rm d}Q(\zeta)} =\sum_{\nu=1}^{d_1+1+2g}\hspace{-6pt}\res{ \zeta =
  p_\nu}\frac{\omega_i(\zeta){\rm d}S(\zeta)^2}{{\rm
    d}P(\zeta){\rm d}Q(\zeta)} 
\eea
The third derivative w.r.t. $t$ has to be treated with some care,
although the result is a formula of the same sort.
For the purpose of this computation it is best to choose as basepoint
defining $\lambda$ the marked point $\infty_Q$, regularizing
appropriately the integral 
\be
\ln\lambda = \int_{\infty_Q}\hspace{-19pt}\setminus{\rm d}S :=
\lim_{\zeta\to\infty_Q} \int_\zeta{\rm d}S - \ln(Q(\zeta))
=\lim_{\zeta\to\infty_Q} \int_\zeta{\rm d}S - \frac
1{d_1}\ln(P(\zeta)) + \frac 1{d_1}\ln(v_{d_2+1})\ .  
\ee
This is just a convenient normalization of the function $\lambda$ with
the property that 
\bea
(\pa_t\ln(\lambda))_P = \int_{\infty_Q}\!\!\!\!(\pa_t{\rm d}S)_P\ ,\qquad
(\pa_t\ln(\lambda))_Q = \int_{\infty_Q}\!\!\!\!(\pa_t{\rm d}S)_Q\ .
\eea
% This
%is so because under variation at $P$ or $Q$ fixed the marked points
%(which are poles of both) do not undergo a variation. Thus we will use
%now $\lambda = \exp(\int_{\infty_Q} {\rm d}S)$; more precisely (since
%this integral should be understood properly regularized) we simply mean that
%we normalize $\lambda$ by the requirement ${\rm d}\lambda/{\rm
%d}Q|_{\infty_Q} = 1$ on the fundamental cell of the universal cover.
Using these properties we find
\bea
\pa_{t}^3 \mathcal F_{g}&& = \pa_t\ln(\gamma\wt\gamma) =
\pa_t\le(\res{\infty_Q}\ln(Q/\lambda)\frac {{\rm d}\lambda}\lambda -\res{\infty_P}\ln(P
\lambda)\frac {{\rm d}\lambda}\lambda 
\ri)=\\
&&=\res{\infty_Q} \frac {(\pa_t Q)_\lambda {\rm d}\lambda}{Q\lambda} -
\res{\infty_P} \frac {(\pa_tP)_\lambda {\rm d}\lambda}{ P\lambda} = \\
&&=-\res{\infty_Q} \frac {(\pa_t \lambda)_Q {\rm d}Q}{Q\lambda} +
\res{\infty_P} \frac {(\pa_t\lambda)_P{\rm d}P}{ P\lambda} = \\
&&= -\res{\infty_Q} \frac {(\pa_t \ln(\lambda))_Q {\rm d}Q}{Q} +
\res{\infty_P} \frac {(\pa_t\ln(\lambda))_P{\rm d}P}{ P}=\\
&&= -\res{\infty_Q} \frac { {\rm d}Q\int_{\infty_Q}\int_{\infty_Q}^{\infty_P} (\pa_t \Omega)_Q}{Q} +
\res{\infty_P}\frac{ {\rm d}P\int_{\infty_Q}\int_{\infty_Q}^{\infty_P} (\pa_t
  \Omega)_P}{ P}=\\
&&= \overbrace{\int_{\infty_Q}^{\infty_Q}\int_{\infty_P}^{\infty_Q} (\pa_t \Omega)_Q}^{=0}
-\int_{\infty_Q}^{\infty_P}\int_{\infty_Q}^{\infty_P} (\pa_t \Omega)_P= \\
&&= \sum_{\mu=1}^{d_2+1+2g}\hspace{-6pt}\res{ \zeta = q_\mu}\int_{\zeta_1=\infty_Q}^{\infty_P}\int_{\zeta_2=\infty_Q}^{\infty_P}
 \int_{\zeta_3=\infty_Q}^{\infty_P}
\Omega^{(3,1)}(\zeta_1,\zeta_2,\zeta_3;\zeta) = \\
&&=  \sum_{\mu=1}^{d_2+1+2g}\hspace{-6pt}\res{ \zeta =
  q_\mu}\frac{{\rm d}S(\zeta)^3}{{\rm
    d}P(\zeta){\rm d}Q(\zeta)} =-\sum_{\nu=1}^{d_1+1+2g}\hspace{-6pt}\res{ \zeta =
  p_\nu}\frac{{\rm d}S(\zeta)^3}{{\rm
    d}P(\zeta){\rm d}Q(\zeta)} \ .
\eea
The other derivatives are
\bea
\pa_{\epsilon_i}\pa_{\epsilon_j}\pa_{u_K} \mathcal F_g &=&-
\sum_{\mu=1}^{d_2+1+2g} \hspace{-6pt}\res{\zeta=q_\mu} \frac
    {\omega_j(\zeta)\omega_k(\zeta) \mathcal U_K(\zeta)}{{\rm
	d}P(\zeta){\rm d}Q(\zeta)} = \sum_{\nu=1}^{d_1+1+2g} \hspace{-6pt}\res{\zeta=p_\nu} \frac
    {\omega_j(\zeta)\omega_k(\zeta) \mathcal U_K(\zeta)}{{\rm
	d}P(\zeta){\rm d}Q(\zeta)}\\
\pa_{\epsilon_i}\pa_{\epsilon_j}\pa_{v_J} \mathcal F_g &=&
-\sum_{\mu=1}^{d_2+1+2g} \hspace{-6pt}\res{\zeta=q_\mu} \frac
    {\omega_j(\zeta)\omega_k(\zeta) \mathcal V_J(\zeta)}{{\rm
	d}P(\zeta){\rm d}Q(\zeta)} =\sum_{\nu=1}^{d_1+1+2g} \hspace{-6pt}\res{\zeta=p_\nu} \frac
    {\omega_j(\zeta)\omega_k(\zeta) \mathcal V_J(\zeta)}{{\rm
	d}P(\zeta){\rm d}Q(\zeta)}\\
\pa_{\epsilon_i}\pa_{u_K} \pa_{v_J} \mathcal F_g &=&
-\sum_{\mu=1}^{d_2+1+2g} \hspace{-6pt}\res{\zeta=q_\mu} \frac
    {\omega_j(\zeta)\mathcal U_K(\zeta)  \mathcal V_J(\zeta)}{{\rm
	d}P(\zeta){\rm d}Q(\zeta)} = \sum_{\nu=1}^{d_1+1+2g} \hspace{-6pt}\res{\zeta=p_\nu} \frac
    {\omega_j(\zeta)\mathcal U_K(\zeta) \mathcal V_J(\zeta)}{{\rm
	d}P(\zeta){\rm d}Q(\zeta)}\\
\pa_t\pa_{u_K}\pa_{v_J} \mathcal F_g &=& \sum_{\mu=1}^{d_2+1+2g} \hspace{-6pt}\res{\zeta=q_\mu} \frac
    {{\rm d}S(\zeta)\mathcal U_K(\zeta)  \mathcal V_J(\zeta)}{{\rm
	d}P(\zeta){\rm d}Q(\zeta)} = -\sum_{\nu=1}^{d_1+1+2g} \hspace{-6pt}\res{\zeta=p_\nu} \frac
    {{\rm d}S(\zeta)\mathcal U_K(\zeta) \mathcal V_J(\zeta)}{{\rm
	d}P(\zeta){\rm d}Q(\zeta)}
\eea
In all of the above formulas we can move the sum over the residues at
the critical points of either maps $P$ and $Q$ because the
differential has poles only  at those points and not at the marked points. Indeed the
denominator has a pole of order $d_1+3$ and $d_2+3$ at $\infty_Q$ and
$\infty_P$ respectively, and the pole of the numerator never exceeds
these values.
In the remaining derivatives, instead, the simplest form is obtained
by summing over only the critical points of one of the two maps,
case by case:
\bea
\pa_t^2\pa_{u_K} \mathcal F_g &=& -\!\!\!\sum_{\mu=1}^{d_2+1+2g} \hspace{-6pt}\res{\zeta=q_\mu} \frac
    {{\rm d}S(\zeta)^2\mathcal U_K(\zeta) }{{\rm
	d}P(\zeta){\rm d}Q(\zeta)}
\\
\pa_t^2\pa_{v_J} \mathcal F_g
 &=& \sum_{\nu=1}^{d_1+1+2g} \hspace{-6pt}\res{\zeta=p_\nu} \frac
    {{\rm d}S(\zeta)^2 \mathcal V_J(\zeta)}{{\rm
	d}P(\zeta){\rm d}Q(\zeta)}
\eea
\bea
\pa_t\pa_{u_K}\pa_{u_J} \mathcal F_g &=& \sum_{\mu=1}^{d_2+1+2g} \hspace{-6pt}\res{\zeta=q_\mu} \frac
    {{\rm d}S(\zeta)\mathcal U_K(\zeta)  \mathcal U_J(\zeta)}{{\rm
	d}P(\zeta){\rm d}Q(\zeta)}\\
\pa_{\epsilon_i}\pa_{u_K}\pa_{u_J} \mathcal F_g &=&-\!\!\! \sum_{\mu=1}^{d_2+1+2g} \hspace{-6pt}\res{\zeta=q_\mu} \frac
    {\omega_i(\zeta)\mathcal U_K(\zeta)  \mathcal U_J(\zeta)}{{\rm
	d}P(\zeta){\rm d}Q(\zeta)}\\
\pa_{u_L} \pa_{u_K}\pa_{u_J} \mathcal F_g &=& -\!\!\!\sum_{\mu=1}^{d_2+1+2g} \hspace{-6pt}\res{\zeta=q_\mu} \frac
    {\mathcal U_L(\zeta)\mathcal U_K(\zeta)  \mathcal U_J(\zeta)}{{\rm
	d}P(\zeta){\rm d}Q(\zeta)}\\
\pa_{v_L} \pa_{u_K}\pa_{u_J} \mathcal F_g &=&-\!\!\! \sum_{\mu=1}^{d_2+1+2g} \hspace{-6pt}\res{\zeta=q_\mu} \frac
    {\mathcal V_L(\zeta)\mathcal U_K(\zeta)  \mathcal U_J(\zeta)}{{\rm
	d}P(\zeta){\rm d}Q(\zeta)}
\eea
\bea
\pa_t\pa_{v_K}\pa_{v_J} \mathcal F_g &=& -\!\!\!\!\sum_{\nu=1}^{d_1+1+2g} \hspace{-6pt}\res{\zeta=p_\nu} \frac
    {{\rm d}S(\zeta)\mathcal V_K(\zeta)  \mathcal V_J(\zeta)}{{\rm
	d}P(\zeta){\rm d}Q(\zeta)}\\
\pa_{\epsilon_i}\pa_{v_K}\pa_{v_J} \mathcal F_g &=& \sum_{\nu=1}^{d_1+1+2g} \hspace{-6pt}\res{\zeta=p_\nu} \frac
    {\omega_i(\zeta)\mathcal V_K(\zeta)  \mathcal V_J(\zeta)}{{\rm
	d}P(\zeta){\rm d}Q(\zeta)}\\
\pa_{v_L}\pa_{v_K}\pa_{v_J} \mathcal F_g &=& \sum_{\nu=1}^{d_1+1+2g} \hspace{-6pt}\res{\zeta=p_\nu} \frac
    {\mathcal V_L(\zeta)\mathcal V_K(\zeta)  \mathcal V_J(\zeta)}{{\rm
	d}P(\zeta){\rm d}Q(\zeta)}\\
\pa_{u_L}\pa_{v_K}\pa_{v_J} \mathcal F_g &=& \sum_{\nu=1}^{d_1+1+2g} \hspace{-6pt}\res{\zeta=p_\nu} \frac
    {\mathcal U_L(\zeta)\mathcal V_K(\zeta)  \mathcal V_J(\zeta)}{{\rm
	d}P(\zeta){\rm d}Q(\zeta)}\ .
\eea
Although we have not written it each time, it is clear that any of
these derivatives is obtained by means of suitable integrals of the
kernel $\Omega^{(3,1)}(\bullet,\bullet,\bullet;\zeta)$  over
\begin{enumerate}
\item a $b$-cycle corresponding to each variation w.r.t. $\epsilon$'s
  (times $\frac 1{2i\pi}$);
\item a circle around $\infty_Q$ against  $Q^K/(2i\pi\,K)$ for
  each variation w.r.t. $u_K$'s;
\item a circle around $\infty_P$ against  $-P^J/(2i\pi \, J)$ for
  each variation w.r.t. $v_J$'s;
\item a path from $\infty_P$ to $\infty_Q$ for each variation
  w.r.t. $t$;
\end{enumerate}
followed by minus the sum over the residues at the points $\zeta= q_\mu$ if there are two
or more derivatives w.r.t the $u_K$'s or two derivatives w.r.t. $t$
and one w.r.t. $u_K$, \\
 or \\
 the sum of residues at
the points $\zeta=p_\nu$'s if there are two or more derivatives w.r.t. the
$v_J$'s or two derivatives w.r.t. $t$
and one w.r.t. $v_J$. 
 In all other cases the choice of residues is immaterial.\par\vskip 4pt
Using the above results somewhat liberally in the case of potentials
given by power series we obtain
\bc[Three--loops correlators]
Denoting by $\zeta(x)$ the inverse function of $Q(\zeta)=x$ 
and $\wt \zeta(y)$ the inverse function of $P(\zeta)=y$
on the respective physical sheets we can write the three--loop
correlators as follows
\bea
\le\langle \tr\frac 1 {x-M_1}\tr \frac 1{x'-M_1} \tr \frac 1
	   {x''-M_1}\ri\rangle_{\hbox{conn}} = -\frac{\ds \sum_{\mu}\res{\zeta=
	     q_\mu}
	   \Omega^{(3,1)}(\zeta(x),\zeta(x'),\zeta(x'');\zeta)}{{\rm
	       d}Q(\zeta(x)){\rm d}Q(\zeta(x')) {\rm
	       d}Q(\zeta(x''))}\\
\le\langle \tr\frac 1 {x-M_1}\tr \frac 1{x'-M_1} \tr \frac 1
	   {y-M_2}\ri\rangle_{\hbox{conn}} = -\frac{\ds \sum_{\mu}\res{\zeta=
	     q_\mu}
	   \Omega^{(3,1)}(\zeta(x),\zeta(x'),\wt\zeta(y);\zeta)}{{\rm
	       d}Q(\zeta(x)){\rm d}Q(\zeta(x')) {\rm
	       d}P(\wt\zeta(x''))}\ ,
\eea
and similarly for the other two three--loop correlators but summing
over $p_\nu$'s and with an overall minus sign.
\ec
{\bf ``Proof''.}
The  quotes are because a complete proof should address also the issue
of 
convergence of the series involved. Indeed we can obtain the above
correlators by applying the relevant puncture operators, which can be
done only by considering infinite series for the potentials. At the
end of this process we should set all the coefficients but a finite
number to zero so that the final formula involves only a finite sum of
residues. Since we are not addressing the summability of the series
involved in this procedure we cannot claim rigor for this proof.
With this in mind we nevertheless proceed on a formal level. For
instance the first correlator would be 
\bea
\sum_{K,K',K''=1}^{\infty}
x^{-K-1}x'^{-K'-1}x''^{-K''-1}KK'K''\pa_{u_K}\pa_{u_{K'}}\pa_{u_{K''}}
\mathcal F_g = \\
=\frac {-1}{(2i\pi)^3}\sum_{K,K',K''=1}^{\infty}\oint_{\infty_Q}\oint_{\infty_Q}
\oint_{\infty_Q} \sum_{\mu}\res{\zeta=q_\mu} \frac{Q(\zeta)^K}{x^K}
\frac {Q(\zeta')^{K'}}{(x')^{K'+1}} \frac
      {Q(\zeta'')^{K''}}{(x'')^{K''+1}}
      \Omega^{(3,1)}(\zeta,\zeta',\zeta'';\zeta) = (\star)\ .
\eea
We can interchange the sums with the integrals (and sum the resulting
geometric series) if the loops around $\infty_Q$ project -e.g.- to
circles in the $Q$--plane which leave $x,x',x''$ in the unbounded
region, that is, if $|Q(\zeta)/x|$, etc. remain bounded and less than
one. In this case we have then
\bea
(\star) = \frac {-1}{(2i\pi)^3}\oint_{\infty_Q}\oint_{\infty_Q}
\oint_{\infty_Q} \sum_{\mu}\res{\zeta=q_\mu} \frac 1{x-Q(\zeta)}
\frac 1{x'-Q(\zeta')} \frac 1
      {x''- Q(\zeta'')}
      \Omega^{(3,1)}(\zeta,\zeta',\zeta'';\zeta) 
\eea
which provides the result by residue evaluation (the only residues are
at the three points $\zeta(x),\zeta(x'),\zeta(x'')$). The other
correlators are computed  exactly in the same way.
``Q. E. D.''\par\vskip 4pt
\section{Conclusion and outlook}
The main result of the paper is the formula for the triple derivatives
of the free energy for arbitrary genus spectral curves. En route we
have computed the variation of the conformal structure of the curve,
which would in principle allow to compute derivatives of any
order. Quite clearly the task becomes an exercise in complication (see
similar problems in \cite{WZ2}). We also remark that in many practical
uses the third derivatives suffices to determine many relevant
properties. In particular our formulas should be sufficient to address
completely the issue of associativity in both sets of variables
$\{u_K\}$ and $\{v_J\}$.
It should be remarked that the multi--loop correlators of the
one-matrix model (which is a sub-case of the two-matrix model when  one
potential is 
Gaussian) were presented by means of a recursive (and hence
not totally explicit) procedure in \cite{ambi}.
%which seems to elude an expression in terms of canonical objects  of
%the curve. I am referring to the mixed correlator 
%\be
%\mathfrak M(x,y):= \le\langle\tr \frac 1 {x-M_1} \frac 1{y-M_2}\ri\rangle\ ,\label{mixedup}
%\ee 
%which cannot be expressed immediately in terms of the free
%energy. It is shown in \cite{eynard2, Bertrand_unpublished} using the
%loop equations that the related (polynomial) mixed correlator
%\be
%G(x,y) = \le\langle\tr \frac {V_1'(x)-V_1'(M_1)} {x-M_1} \frac
%	      {V_2'(y)-V_2'(M_2)} {y-M_2}\ri\rangle\ ,\label{P0}
%\ee
%is precisely what {\em defines} the spectral curve as an algebraic
%equation between $P$ and $Q$ as follows (or vice-versa, the algebraic
%relation between $P$ and $Q$ defines (\ref{P0})),
%\be
%\mathcal E(Q,P):= (V_1'(Q)-P)(V_2'(P)-Q) - G(Q,P) + t \equiv 0\ .
%\ee

We have not considered the mixed correlator 
\be
\mathfrak M(x,y):= \le\langle\tr \frac 1 {x-M_1} \frac 1{y-M_2}\ri\rangle\ .\label{mixedup}
\ee
Such correlator is not universal and can be obtained via loop
equations in the large $N$-limit \cite{staudacher, eynardchain, eynard2,
  chaineloop, eynard}.
The mixed correlator (\ref{mixedup}) has also already been
computed exactly in the finite $N=\dim(M_i)$ regime in \cite{BE}.\par\vskip 6pt
{\bf \large Acknowledgments}. I would like to thank Bertrand Eynard,
John Harnad, Jacques Hurtubise,  Alexey Kokotov,  Dmitri Korotkin,
Alexander Bobenko for
stimulating and critical discussions.
\appendix
\section{Example: the Gaussian--Gaussian model}
If both potentials are Gaussian ($d_1=d_2=1$) then the spectral curve
is {\em a fortiori} rational. There is nothing new in what we will
write in this appendix since the tau--function has already been
computed (see \cite{KKWZ}, with the warning that we are using
different normalizations for the parameters)
corresponding to the elliptic case (it suffices to consider the
holomorphic and antiholomorphic coordinates as independent in their
formulas to obtain ours). We choose
the normalization of the coordinate $\lambda$ so that 
\bea
&&Q(\lambda) = \gamma\lambda + \alpha_0 + \alpha_1 \lambda^{-1}\\
&&P(\lambda) = \gamma\lambda^{-1} + \beta_0 + \beta_1\lambda\ ,\qquad
\gamma \in \R_+\ .
\eea
Using the formula for the free energy and computing the residues one
obtains
\be
2\mathcal F_0 = t^2\ln\le(\frac T{\sqrt{u_2v_2-1}}\ri) - \frac 3 2 t^2
+ t\frac {(u_2{v_1}^2 + v_2u_1^2+ 2 v_1u_1)}{u_2v_2-1}\ .
\ee
In this case one can check by direct computations that our residue
formulas yield correctly the third derivatives of $\mathcal F_0$ (a
computer aided algebra system like Maple will help for the
computations of residues).
\section{Relation with conformal maps}
\label{strnzn} 
We want here to show that the ``residue'' formulas in \cite{WZ2, MWZ1} are
contained in our formulas.
We recall that the comparison is achieved by taking (see Section
\ref{setting}) $z=Q$, $d_1=d_2=d$, $V_1 = \overline{V_2} = V$  and using 
\bea
&& Q(\lambda) = \gamma\lambda + \sum_{i=0}^{d} \alpha_j\lambda^{-i}\\
&& P(\lambda) = \frac \gamma \lambda + \sum_{i=0}^{d}
\ov\alpha_i\lambda^i\\
&&\hspace{1cm}\gamma\in \R_+.
\eea
In this case the Bergman kernel is simply
\be
\Omega(\lambda,\lambda') = \frac {{\rm d}\lambda {\rm
    d}\lambda'}{(\lambda-\lambda')^2}\ ,
\ee
and the differentials of the second kind reduce to 
\be
\mathcal U_K = (\pa_{u_K} P)_Q{\rm d}Q = \frac 1 K \res{\lambda'=\infty}
\Omega(\lambda,\lambda') Q(\lambda)^K = \frac {{\rm d}}{{\rm
    d}\lambda}\le(Q^K(\lambda)\ri)_+{{\rm d}\lambda} =\frac 1 K {\rm
  d} \le(Q^{K}(\lambda)\ri)_+\ ,
\ee
where the subscript $()_+$ means the nonnegative part in $\lambda$.
Then our formulas reproduce exactly those in \cite{MWZ1} when proper
identifications are made (our normalization differs from theirs).
\section{Example: the Gaussian--Cubic case}
Here we show how to compute the free energy and derivatives for the
case $V_2(y) = \frac {v_2}2 y^2 + v_1 y$ and $V_1$ cubic (one
could do the exercise for higher degree potentials, the expressions
become quite large but can be obtained explicitly e.g. with Maple). This case
corresponds to the one--matrix model\footnote{Quite clearly the
  corresponding matrix model would have to be defined with normal
  matrices with spectrum on certain contours, rather than Hermitian
  matrices \cite{BHE1}}  and indeed the spectral curve is
hyperelliptic. We will however restrict ourselves to the genus zero
case where computations can be made completely explicit.
Indeed we have
\bea
Q(\lambda) = \gamma\lambda + \alpha_0 + \frac {\alpha_1}\lambda\\
P(\lambda) = \frac \gamma\lambda + \sum_{j=0}^{2} \beta_j\lambda^j\ .
\eea
We see that here the computations of all derivatives are possible in
explicit terms because $Q$ has only two critical points; in all cases
in which (when computing the third derivatives) we should sum over the
residues at the critical points of $P$ we can replace the sum by
summing over the critical points of $Q$ plus the other residue at
$\lambda = 0$.
 We can eliminate the parameters $\alpha_i$ as follows (sometimes that makes
formulas shorter) 
\be
\alpha_0 = v_1+v_2\beta_0,\ \ \ \alpha_1 = v_2\gamma\ ,\ \ t=
\alpha_1\beta_1 -\gamma^2\ .
\ee
In order to express the free energy in terms of the coordinates $u_k$
we should invert an algebraic set of equations which can be done
numerically at least. Nonetheless we can compute
the free energy and any derivative up to the third using our formulas.
The quadri differential $\Omega^{3,1}$ is simply
\be
\Omega^{(3,1)}(\lambda_1,\lambda_2,\lambda_3;\lambda)  = \frac
{{\rm d}\lambda_1 {\rm d}\lambda_2{\rm d}\lambda_3 {{\rm d}\lambda}
}{ (\lambda_1-\lambda)^2(\lambda_2-\lambda)^2(\lambda_3-\lambda)^2 P'(\lambda)Q'(\lambda)}
\ee
Using e.g. Maple one can immediately compute all derivatives by
residue evaluations.
For instance the chemical potential is 
\bea
\mu =\pa_t\mathcal F_0 ={{\gamma}}^{2}-\alpha_{{0}}\beta_{{0}}-\beta_{{1}}\alpha_{{1}}+2\,t
\ln  \left( {\gamma} \right) +{\frac {4\,\beta_{{2}}\alpha_{{0}}
\alpha_{{1}}+{\alpha_{{0}}}^{2}\beta_{{1}}+{\beta_{{0}}}^{2}
\alpha_{{1}}}{{2\gamma}}}-{\frac {\beta_{{2}}{\alpha_{{0}}}^{3}
}{{3{\gamma}}^{2}}}
\eea
and we have (just two examples)
\bea
&&\hspace{-2cm}\pa_t^3 \mathcal F_0 = {\frac
  {-{{\gamma}}^{3}+v_{{2}}{{\gamma}}^{2}\beta_{{1}}}{{{\gamma}}^{5} -
    4\,{\beta_{{2}}}^{2}{v_{{2}}}^{3}{{\gamma}}^{3}+{{\gamma}}^{3}{v_{{2}}}^{2}{\beta_{{1}}}^{2} - 2\,{{\gamma}}^{4}v_{{2}}  
\beta_{{1}}}}\\
&&\hspace{-2cm}\pa_{v_2}^3\mathcal F_0 =
8\, \bigg(\gamma\, \bigg( {\gamma}^{2}{\beta_{{1}}}^{3}+4\,
{\gamma}^{3}{\beta_{{2}}}^{2}+12\,{\gamma}^{2}\beta_{{0}}
\beta_{{1}}\beta_{{2}}+2\,\gamma\,{\beta_{{0}}}^{3}\beta_{{2}}+3
\,\gamma\,{\beta_{{1}}}^{2}{\beta_{{0}}}^{2}+24\,{\beta_{{2}}}^{3
}\gamma\,\beta_{{0}}{v_{{2}}}^{2} +4\,{\beta_{{1}}}^{2}{\beta_{{2}
}}^{2}\gamma\,{v_{{2}}}^{2}+
\\
&&+12\,\beta_{{1}}{\beta_{{2}}}^{2}{
\beta_{{0}}}^{2}{v_{{2}}}^{2}+4\,{\gamma}^{2}\beta_{{1}}{\beta_{{
2}}}^{2}v_{{2}}+12\,\gamma\,{\beta_{{0}}}^{2}{\beta_{{2}}}^{2}v_{
{2}}-6\,\gamma\,\beta_{{2}}\beta_{{0}}{\beta_{{1}}}^{2}v_{{2}}-{\gamma}\,{\beta_{{1}}}^{4}v_{{2}}-3\,{\beta_{{1}}}^{3}{\beta_{{0}}}
^{2}v_{{2}} \bigg) \bigg)\\
&&/
\bigg(-{\gamma}^{2}+4\,{\beta_{{2}}}^{2}{v_{{2}}}
^{3}-{\beta_{{1}}}^{2}{v_{{2}}}^{2}+2\,v_{{2}}\gamma\,\beta_{{1}}
\bigg) \ .
\eea


\begin{thebibliography}{00}

\bibitem{AvM1} M. Adler and P. Van Moerbeke, ``String-orthogonal polynomials,
string equations and 2-Toda symmetries'', {\it Comm. Pure and Appl. Math. J.},
{\bf 50} (1997), 241-290.

\bibitem{AvM2} M. Adler and P. Van Moerbeke, ``The Spectrum of Coupled Random
Matrices'', {\it Ann. Math.} {\bf 149} (1999), 921--976.

\bibitem{ambi}  J. Ambj\/orn, J. Jurkiewicz, Yu. M. Makeenko,
  ``Multiloop correlators for two--dimensional quantum gravity'',
  Phys. Lett. B {\bf 251}, no. 4 (1990), 517--524 .
%\bibitem{BDJ} J. Baik, P. Deift and K. Johansson ``The Longest increasing
%subsequece in a random permutation and a unitary random matrix model'', {\it J.
%Amer. Math. Soc.} {\bf 12}, 1119-1178 (1999).

\bibitem{DVV} T. Banks, W. Fischler, S. H. Shenker, L. Susskind, ``M theory as 
a matrix model: A conjecture'', {\em Phys. Rev.} {\bf D55}, (1997)
5112.

\bibitem{F1} M. Bertola, ``Free Energy of the Two--Matrix Model/dToda
  Tau--Function'',  Nucl. Phys. B {\bf 669} [FS] (2003), 435--461
\bibitem{Berto} M. Bertola, ``Bilinear semi--classical moment functionals and 
their  integral representation'', J. App. Theory {\bf 121} (2003), 71--99.

\bibitem{BHE1} M. Bertola, B. Eynard, J. Harnad, ``Differential
  systems for biorthogonal polynomials appearing in 2--matrix models
  and the associated Riemann--Hilbert problem'', nlin.SI/0208002, to
  appear in Comm. Math. Phys.

\bibitem{BHE2}  M. Bertola, B. Eynard, J. Harnad, ``Duality of spectral curves 
arising in two--matrix
  models'', Theor. Math. Phys. {\bf 134} (1) (2003), 25--36.

\bibitem{BHE3}  M. Bertola, B. Eynard, J. Harnad, ``Duality, Biorthogonal 
Polynomials and Multi--Matrix
  Models'', Commun. Math. Phys. {\bf 229} (2002), 73--120.

\bibitem {BE} M. Bertola, B. Eynard, ``Mixed Correlation Functions of
the Two-Matrix Model'', J. Phys. A {\bf 36} (2003) 7733--7750.

\bibitem{BI} P.M. Bleher and A.R. Its, eds.,  ``Random Matrix Models and
Their Applications'', MSRI Research Publications {\bf 40}, Cambridge Univ.
Press, (Cambridge, 2001). 

\bibitem{daul} J. M. Daul, V. Kazakov, I. Kostov, ``Rational Theories
  of 2D Gravity from the Two--Matrix Model'', { Nucl. Phys. B} {\bf
  409} (1993), 311--338.

\bibitem{Matrixsurf} F. David, ``Planar diagrams, two-dimensional lattice 
gravity and surface models'', { Nucl. Phys. B}, {\bf  257 [FS14]}
(1985), 45.

\bibitem{ZJDFG} P. Di Francesco, P. Ginsparg, J. Zinn-Justin, ``2D Gravity and 
Random Matrices'', { Phys. Rep.} {\bf 254} (1995), 1.

\bibitem{DKMVZ} P. Deift, T. Kriecherbauer, K. T. R. McLaughlin,
  S. Venakides, Z. Zhou, ``Strong asymptotics of orthogonal
  polynomials with respect to exponential weights'', Comm. Pure
  Appl. Math. {\bf 52} (1999), 1491--1552.

\bibitem{eynard} B. Eynard, ``Eigenvalue distribution of large random matrices,
from one matrix to several coupled matrices'' { Nucl. Phys. B } {\bf 506} (1997), 
633.

\bibitem{eynard2} B. Eynard, ``Large $N$ expansion of the 2-matrix
  model'', {J. High Energy Phys.}   no. 1, {\bf 051} (2003), 38 pp.

\bibitem{Bertrand_unpublished} B. Eynard, ``Large N expansion of the
  2-matrix model, multicut case'', math-ph/0307052.

\bibitem{chaineloop} B. Eynard, ``Master loop equations, free energy
  and correlations for the chain of matrices'', hep-th/0309036.
 
\bibitem{eynardchain} B. Eynard, ``Correlation functions of eigenvalues of 
multi-matrix models, and the limit of a time dependent matrix'', 
{\em J. Phys. A: Math. Gen.} {\bf 31} (1998), 8081.

\bibitem{ercol} N. M. Ercolani and K. T.-R. McLaughlin, presentation
  at the Montr\'eal 2002 AMS meeting.

\bibitem{Fay} J. D. Fay, ``Kernel functions, analytic torsion, and
  moduli spaces'', Memoirs of AMS, 1992, v. 96, n. 464.

\bibitem{Guhr} T. Guhr, A. Mueller-Groeling, H.A. Weidenmuller, ``Random 
matrix theories in quantum physics: Common concepts'', {\em Phys. Rep.}
{\bf 299} (1998), 189.

\bibitem{harnack} A. Harnack, ``\"Uber die Vieltheilgkeit der ebenen
  algebraischen Curven'', Math. Ann. {\bf 10} (1876), 189--198.

\bibitem{Kazakov} V.A. Kazakov, ``Ising model on a dynamical planar random
lattice: exact solution'', {\em Phys Lett.} {\bf A119} (1986), 140-144. 

\bibitem{curve} V. A. Kazakov, A. Marshakov, ``Complex  Curve of the
  Two Matrix Model and its Tau-function'', J. Phys. A  {\bf 36}  (2003),  no. 12, 3107--3136. 

\bibitem{korotkin} A. Kokotov, D. Korotkin, ``Tau-function on Hurwitz
  spaces'', math-ph/0202034

\bibitem{korotkin2} A. Kokotov, D. Korotkin, ``On G--function of
  Frobenius manifolds related to Hurwitz spaces'', math-ph/0306053.

\bibitem {KKWZ} I. K. Kostov, I. Krichever, P. Wiegmann, A. Zabrodin,
  ``The $\tau$-function for analytic curves'', 
{\em Random matrix models and their applications},  Math. Sci. Res. Inst. Publ.
{\bf 40}, 285--299,  Cambridge Univ. Press, Cambridge, 2001.

\bibitem {KMZ} I. Krichever, A. Marshakov, A. Zabrodin, ``Integrable
  Structure of the Dirichlet Boundary Problem in Multiply-Connected
  Domains'', hep-th/0309010. 

\bibitem{Kric} I. Krichever, ``The $\tau$-Function of the Universal
  Whitham Hierarchy, Matrix Models and Topological Field Theories'',
  Comm. Pure Appl. Math. {\bf 47}  (1994),  no. 4, 437--475.

\bibitem {MWZ1} A. Marshakov, P. Wiegmann, A. Zabrodin, ``Integrable
  structure of the Dirichlet boundary problem in two dimensions'',
  Comm. Math. Phys. {\bf 227} no 1. (2002), 131--153.

\bibitem{Mehta} M.L. Mehta, {\em Random Matrices}, 2nd edition, (Academic
Press, New York, 1991).

\bibitem{spohn} M. Praehofer and H. Spohn, ``Universal distributions for
growth  processes in $1+1$ dimensions and random matrices'', {\em Phys. Rev.
Lett.}   {\bf 84} (2000) 4882.

\bibitem{rauch} H. E. Rauch, ``Weierstrass points, branch points, and
  moduli of Riemann surfaces'', Comm. Pure Appl. Math., {\bf 12} (1959),
  543--560.


\bibitem{staudacher} M.  Staudacher, ``Combinatorial solution of the
  two-matrix model'',  Phys. Lett. B  {\bf 305} (1993),  no. 4,
  332--338.
 
\bibitem{tt1} K. Takasaki, T. Takebe, ``SDiff(2) Toda
  equation--hierarchy, tau function, and symmetries'',
  Lett. Math. Phys. {\bf 23} (1991), no.3, 205-214.

\bibitem{tt2}  K. Takasaki, T. Takebe, ``SDiff(2) KP hierarchy,
  Infinite analysis, Part A, B'', (Kyoto, 1991),
  Adv. Ser. Math. Phys., {\bf 16}, World Sci. Publishing, River Edge,
  NJ, 1992, 889--922.

\bibitem{tt3}  K. Takasaki, T. Takebe, ``Quasi--classical limit of
  Toda hierarchy and W-infinity symmetries'', Lett. Math. Phys. {\bf
  28} (1993), no. 3, 165--176.

\bibitem{tt4} K. Takasaki, T. Takebe, ``Integrable Hierarchies and
  Dispersionless Limit'',   Rev. Math. Phys.  {\bf 7}  (1995),  no. 5, 743--808.


\bibitem {Tak} L. A. Takhtajan, ``Free Bosons and Tau-Functions for
   Compact Riemann Surfaces and Closed Smooth Jordan Curves. Current
   Correlation Functions'', Lett. Math. Phys. {\bf 56} (2001), 181--228.

\bibitem {Teo} L-P. Teo, ``Analytic functions and integrable
  hierarchies---characterization of tau functions'',
  Lett. Math. Phys.  {\bf 64}  (2003),  no. 1, 75--92. 

\bibitem{UT} K. Ueno and K. Takasaki, ``Toda Lattice Hierarchy'',
{\it Adv. Studies Pure Math.} {\bf 4} (1984), 1--95.

\bibitem{Verbaarshot} J.J.M. Verbaarshot, ``Random matrix model approach to 
chiral symmetry'', {\em Nucl. Phys. Proc. Suppl.} {\bf 53} (1997), 88.

\bibitem {WZ} P. Wiegmann, A. Zabrodin, ``Conformal maps and integrable
  hierarchies'', {\em Comm. Math. Phys.}, {\bf 213} (2000), no.3, 523--538.

\bibitem{WZ2}  P. Wiegmann, A. Zabrodin, `` Large scale correlations
  in normal and general non-Hermitian matrix ensembles'',  J.Phys. A
  {\bf 36} (2003), 3411-3424.

\bibitem {Zab} A. Zabrodin, ``The dispersionless limit of the Hirota equations in some problems of
   complex analysis'', {\em Teoret. Mat. Fiz.}, {\bf 129} (2001),
   no. 2, 239--257.

\end{thebibliography}
\end{document}